\documentclass[letterpaper]{sig-alternate}
\usepackage{graphicx}

\def\sharedaffiliation{
\end{tabular}
\begin{tabular}{c}}

\begin{document}

\title{What's in a Session:\\Tracking Individual Behavior on the Web}

\numberofauthors{5}
\author{
\alignauthor
  Mark Meiss$^{1,2}$\\
  \email{mmeiss@indiana.edu}
\alignauthor
  John Duncan$^{1}$\\
  \email{johfdunc@indiana.edu}
\alignauthor
  Bruno Gon\c{c}alves$^{1}$\\
  \email{bgoncalves@gmail.com}
\and
\alignauthor
  Jos\'{e} J. Ramasco$^{3}$\\
  \email{jramasco@isi.it}
\alignauthor
  Filippo Menczer$^{1,3}$\\
  \email{fil@indiana.edu}
\sharedaffiliation
\affaddr{$^1$School of Informatics, Indiana University, Bloomington, IN, USA}\\
\affaddr{$^2$Advanced Network Management Lab, Indiana University, Bloomington, IN, USA}\\
\affaddr{$^3$Complex Networks Lagrange Laboratory, ISI Foundation, Torino, Italy}
}

\maketitle

%------------------------------------------------------------------------
% Abstract
%------------------------------------------------------------------------

\begin{abstract}
  We examine the properties of all HTTP requests generated by a
  thousand undergraduates over a span of two months.  Preserving user
  identity in the data set allows us to discover novel properties of
  Web traffic that directly affect models of hypertext navigation.  We
  find that the popularity of Web sites---the number of users who
  contribute to their traffic---lacks any intrinsic mean and may be
  unbounded.  Further, many aspects of the browsing behavior of
  individual users can be approximated by log-normal distributions
  even though their aggregate behavior is scale-free.  Finally, we
  show that users' click streams cannot be cleanly segmented into
  sessions using timeouts, affecting any attempt to model hypertext
  navigation using statistics of individual sessions.  We propose a
  strictly logical definition of sessions based on browsing activity
  as revealed by referrer URLs; a user may have several active
  sessions in their click stream at any one time.  We demonstrate that
  applying a timeout to these logical sessions affects their
  statistics to a lesser extent than a purely timeout-based mechanism.
\end{abstract}

%------------------------------------------------------------------------
% Other front material.
%------------------------------------------------------------------------




%------------------------------------------------------------------------
% (1) Introduction
%------------------------------------------------------------------------

\section{Introduction}

We report our analysis of the Web traffic of approximately one
thousand residential users over a two-month period.  This data set
preserves the distinctions between individual users, making possible
detailed per-user analysis.  We believe this is the largest study to
date to examine the complete click streams of so many users in their
place of residence for an extended period of time, allowing us to
observe how actual users navigate a hyperlinked information space
while not under direct observation.  The first contributions of this
work include the discoveries that the popularity of Web sites as
measured by distinct visitors is unbounded; that many of the power-law
distributions previously observed in Web traffic are aggregates of
log-normal distributions at the user level; and that there exist two
populations of users who are distinguished by whether or not their Web
activity is largely mediated by portal sites.

A second set of contributions concerns our analysis of browsing
sessions within the click streams of individual users.  The concept of
a Web session is critical to modeling real-world navigation of
hypertext, understanding the impact of search engines, developing
techniques to identify automated navigation and retrieval, and
creating means of anonymizing (and de-anonymizing) user activity on
the Web.  We show that a simple timeout-based approach is inadequate
for identifying sessions and present an algorithm for segmenting a
click stream into \textit{logical sessions} based on referrer
information.  We use the properties of these logical sessions to show
that actual users navigate hypertext in ways that violate a stateless
random surfer model and require the addition of backtracking or
branching.

Finally, we emphasize which aspects of this data present possible
opportunities for anomaly detection in Web traffic.  Robust anomaly
detection using these properties makes it possible to uncover ``bots''
masquerading as legitimate user agents.  It may also undermine the
effectiveness of anonymization tools, making it necessary to obscure
additional properties of a user's Web surfing to avoid betraying their
identity.

\subsection*{Contributions and Outline}

In the remainder of this paper, after some background and related
work, we describe the source and collection procedures of our Web
traffic data.  The raw data set includes over 400 million HTTP
requests generated by over a thousand residential users over the
course of two months, and we believe it to provide the most accurate
picture to date of the hypertext browsing behavior of individual users
as observed directly from the network.

Our main contributions are organized into three sections:

\begin{itemize}

 \item{We confirm earlier findings of scale-free distributions for
 various per-site traffic properties aggregated across users.  We show
 this also holds for site popularity as measured by the number of
 unique vistors.  (\S~\ref{section-host})}

 \item{We offer the first characterization of individual traffic
 patterns involving continuous collection from a large population.  We
 find that properties such as jump frequency, browsing rates, and the
 use of portals are not scale-free, but rather log-normally
 distributed.  Only when aggregated across users do these properties
 exhibit scale-free behavior.  (\S~\ref{section-user})}

 \item{We investigate the notion of a Web ``session,'' showing that
 neither a simple timeout nor a rolling average provide a robust
 definition.  We propose an alternative notion of \textit{logical}
 session and provide an algorithm for its construction.  While logical
 sessions have no inherent temporal scale, they are amenable to the
 addition of a timeout with little net effect on their statistical
 properties.  (\S~\ref{section-session})}

\end{itemize}

We conclude with a discussion of the limitations of our data, the
implications of this work for modeling and anomaly detection, and
potential future work in the area.

%------------------------------------------------------------------------
% (2) Background
%------------------------------------------------------------------------

\section{Background}

Internet researchers have been quick to recognize that structural
analysis of the Web becomes far more useful when combined with actual
\emph{behavioral} data.  The link structure of the Web can differ
greatly from the set of paths that are actually navigated, and it
tells us little about the behavior of individual users.  A variety of
behavioral data sources exist that can allow researchers to identify
these paths and improve Web models accordingly.  The earliest efforts
have used browser logs to characterize user navigation
patterns~\cite{catledge95characterizing}, time spent on pages,
bookmark usage, page revisit frequencies, and overlap among user
paths~\cite{cockburn-what}.  The most direct source of behavioral data
comes from the logs of Web servers, which have been used for
applications such as personalization~\cite{mobasher-2000-automatic}
and improving caching behavior~\cite{yang-2003-weblog}.  More recent
efforts involving server logs have met with notable success in
describing typical user behavior~\cite{goncalves08-2}.  Because search
engines serve a central role in users' navigation, their log data is
particularly useful in improving search results based on user
behavior~\cite{agichtein-2006-user,luxenburger-2004-querylog}.

Other researchers have turned to the Internet itself as a source of
data on Web behavior.  Network flow data generated by routers, which
incorporates high-level details of Internet connections without
revealing the contents of individual packets, has been used to
identify statistical properties of Web user behavior and discriminate
peer-to-peer traffic from genuine Web
activity~\cite{Meiss05NetFlow,meiss-2008-structural,erman-2007-identifying}.

The most detailed source of behavioral data consists of actual Web
traffic captured from a running network, as we do here.  The present
study most closely relates to the work of Qiu
\textit{et al.}~\cite{Cho05WebDB}, 
who used captured HTTP packet traces to investigate a variety of
statistical properties of users' browsing behavior, especially the
extent on which they appear to rely on search engines in their
navigation of the Web.

We have also used captured HTTP requests in our previous work to
describe ways in which PageRank's random-surfer model fails to
approximate actual user behavior, which calls into question its use
for ranking search results~\cite{meiss-2008-wsdm}.  One way of
overcoming these shortcomings is to substitute actual traffic data for
ranking pages~\cite{liu-2008-browserank}.  However, this may create a
feedback cycle in which traffic grows super-linearly with popularity,
leading to a situation (sometimes called ``Googlearchy'') in which a
few popular sites dominate the Web and lesser known sites are
difficult to
discover~\cite{PandeyRoyOlstonChoChak05VLDB,Fortunato05egalitarian}.
More importantly for the present work, simply accepting traffic data
as a given does not further our understanding of user behavior.  We
can also overcome the deficiencies of the random-surfer model by
improving the model itself.  This paper offers analysis of key
features of observed behavior to support the development of improved
agent-based models of Web traffic~\cite{goncalves-2008-remembering}.

The present study also relates to work in anomaly detection and
anonymization software for the Web.  The Web Tap project, for example,
attempted to discover anomalous traffic requests using metrics such as
request regularity and interrequest delay time, quantities which we
discuss in the present work~\cite{borders-2004-webtap}.  The success
of systems that aim to preserve the anonymity of Web users is known to
be dependent on a variety of empirical properties of behavioral data,
some of which we directly address here~\cite{viecco-2009-privacy}.

%------------------------------------------------------------------------
% (3) Web click data
%------------------------------------------------------------------------

\section{Data Description}

\subsection{Data Source}

\begin{figure}[tb]
\begin{center}
\includegraphics[width=0.80\columnwidth]{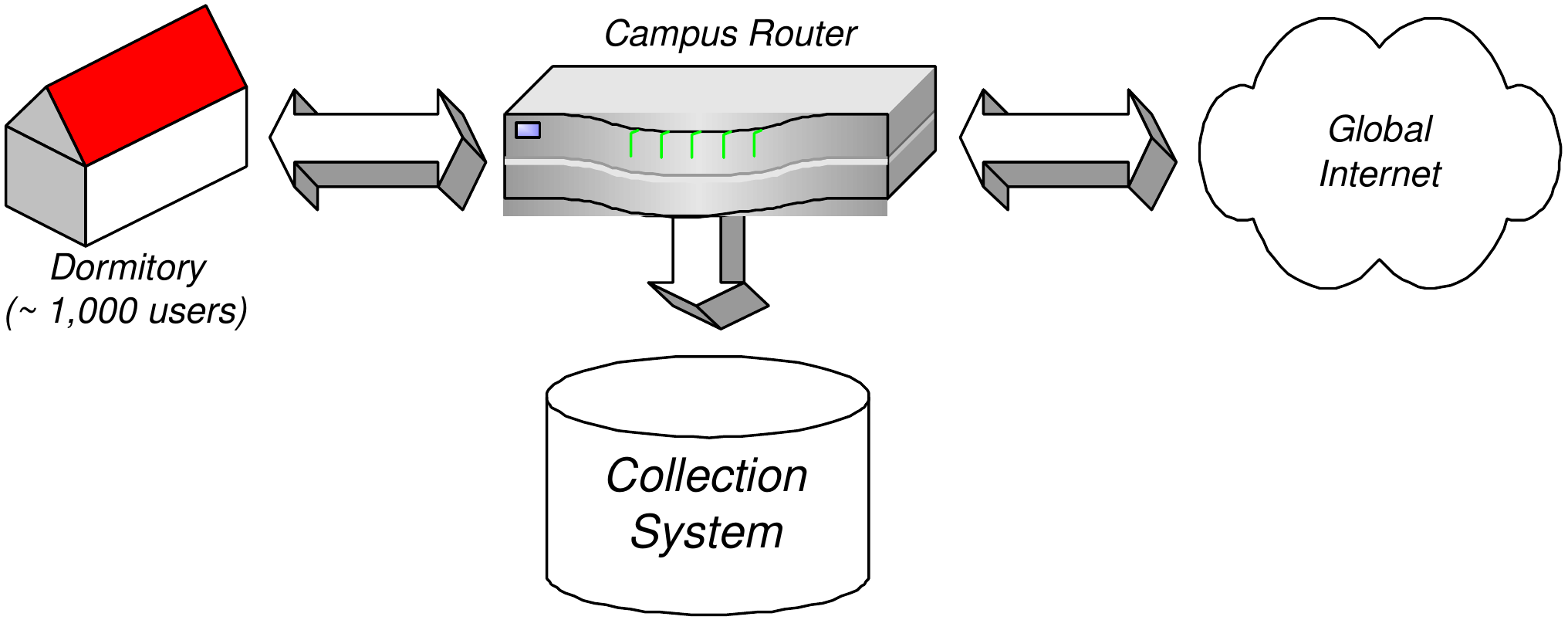}
\end{center}
\caption{
  System architecture for data collection.}
\label{fig:architecture}
\end{figure}

The click data we use in this study was gathered from a dedicated
FreeBSD server located in the central routing facility of the
Bloomington campus of Indiana University
(Figure~\ref{fig:architecture}).  This system had a 1~Gbps Ethernet
port that received a mirror of all outbound network traffic from one
of the undergraduate dormitories.  This dormitory consists of four
wings of five floors each and is home to just over a thousand
undergraduates.  Its population is split roughly evenly between men
and women, and its location causes it to have a somewhat greater
proportion of music and education students than other campus housing.

To obtain information on individual HTTP requests passing over this
interface, we first use a Berkeley Packet Filter to capture only
packets destined for TCP port 80.  While this eliminates from
consideration all Web traffic running on non-standard ports, it does
give us access to the great majority of it.  We make no attempt to
capture or analyze encrypted (HTTPS) traffic using TCP port 443.  Once
we have obtained a packet destined for port 80, we use a regular
expression search against the payload of the packet to determine
whether it contains an HTTP GET request.

If we do find an HTTP GET request in the packet, we analyze the packet
further to determine the virtual host contacted, the path requested,
the referring URL, and the advertised identity of the user agent.  We
then write a record to our raw data files that contains the MAC
address of the client system, a timestamp, the virtual host, the path
requested, the referring URL, and a flag indicating whether the user
agent matches a mainstream browser (Internet Explorer,
Mozilla/Firefox, Safari, or Opera).  We maintain record of the MAC
address only in order to distinguish the traffic of individual users.
We thus assume that most computers in the building have a single
primary user, which is reasonable in light of the connectedness of the
student population (only a small number of public workstations are
available in the dormitory).  Furthermore, as long as the users do not
replace the network interface in their computer, this information
remains constant.

The aggregate traffic of the dormitory was sufficiently low so that
our sniffing system could maintain a full rate of collection without
dropping packets.  While our collection system offers a rare
opportunity to capture the complete browsing activity of a large user
population, we do recognize some potential disadvantages of our data
source.  Because we do not perform TCP stream reassembly, we can only
analyze HTTP requests that fit in a single 1,500 byte Ethernet frame.
While the vast majority of requests do so, some GET-based Web services
generate extremely long URLs.  Without stream reassembly, we cannot
log the Web server's response to each request: some requests will
result in redirections or server errors, and we are unable to
determine which ones.  Finally, a user can spoof the HTTP referrer
field; we assume that few students do so, and those who do generate a
small portion of the overall traffic.

\subsection{Data Dimensions}

The click data was collected over a period of about two months, from
March 5, 2008 through May 3, 2008.  This period included a week-long
vacation during which no students were present in the building.
During the full data collection period, we logged nearly 408 million
HTTP requests from a total of 1,083 unique MAC addresses.

Not every HTTP request from a client is indicative of an actual human
being trying to fetch a Web page; in fact, such requests actually
constitute a minority of all HTTP requests.  For this reason, we
retain only those URLs that are likely to be requests for actual Web
pages, as opposed to media files, style sheets, Javascript code,
images, and so forth.  This determination is based on the extension of
the URL requested, which is imprecise but functions well as a
heuristic in the absence of access to the HTTP \textit{Content-type}
header in the server responses.  We also filtered out a small subset
of users with negligible activity; their traffic consisted largely of
automated Windows Update requests and did not provide meaningful data
about user activity.  Finally, we also discovered the presence of a
poorly-written anonymization service that was attempting to obscure
traffic to a particular adult chat site by spoofing requests from
hundreds of uninvolved clients.  These requests were also removed from
the data set.

We found that some Web clients issue duplicate HTTP requests (same
referring URL and same target URL) in nearly simultaneous bursts.
These bursts occur independently of the type of URL being requested
and are less than a single second wide.  We conjecture that they may
involve checking for updated content, but we are unable to confirm
this without access to the original HTTP headers.  Because this
behavior is so rapid that it cannot reflect deliberate activity of
individual users, we also removed the duplicate requests from the data
set.

Privacy concerns and our agreement with the Human Subjects Committee
of our institution also obliged us to try to remove all identifying
information from the referring and target URLs.  One means of doing so
is to strip off all identifiable query parameters from the URLs.
Applying this anonymization procedure affects roughly one-third of the
remaining requests.

The resulting data set (summarized in Table~\ref{table:data}) is the
basis for all of the description and analysis that follows.

\begin{table}
\caption{Approximate dimensions of the filtered and anonymized data set.}
\begin{center}
\begin{tabular}{|l|c|}
\hline
Page requests & 29.8 million \\
Unique users & 967 \\
Web servers & 630,000 \\
Referring hosts & 110,000 \\
\hline
\end{tabular}
\end{center}
\label{table:data}
\end{table}

%------------------------------------------------------------------------
% (4) Host-based properties
%------------------------------------------------------------------------

\section{Host-based properties}
\label{section-host}

\begin{figure}[tb]
\begin{center}
\begin{tabular}[t]{cc}
\raisebox{1.8in}{A} & \includegraphics[width=0.75\columnwidth]{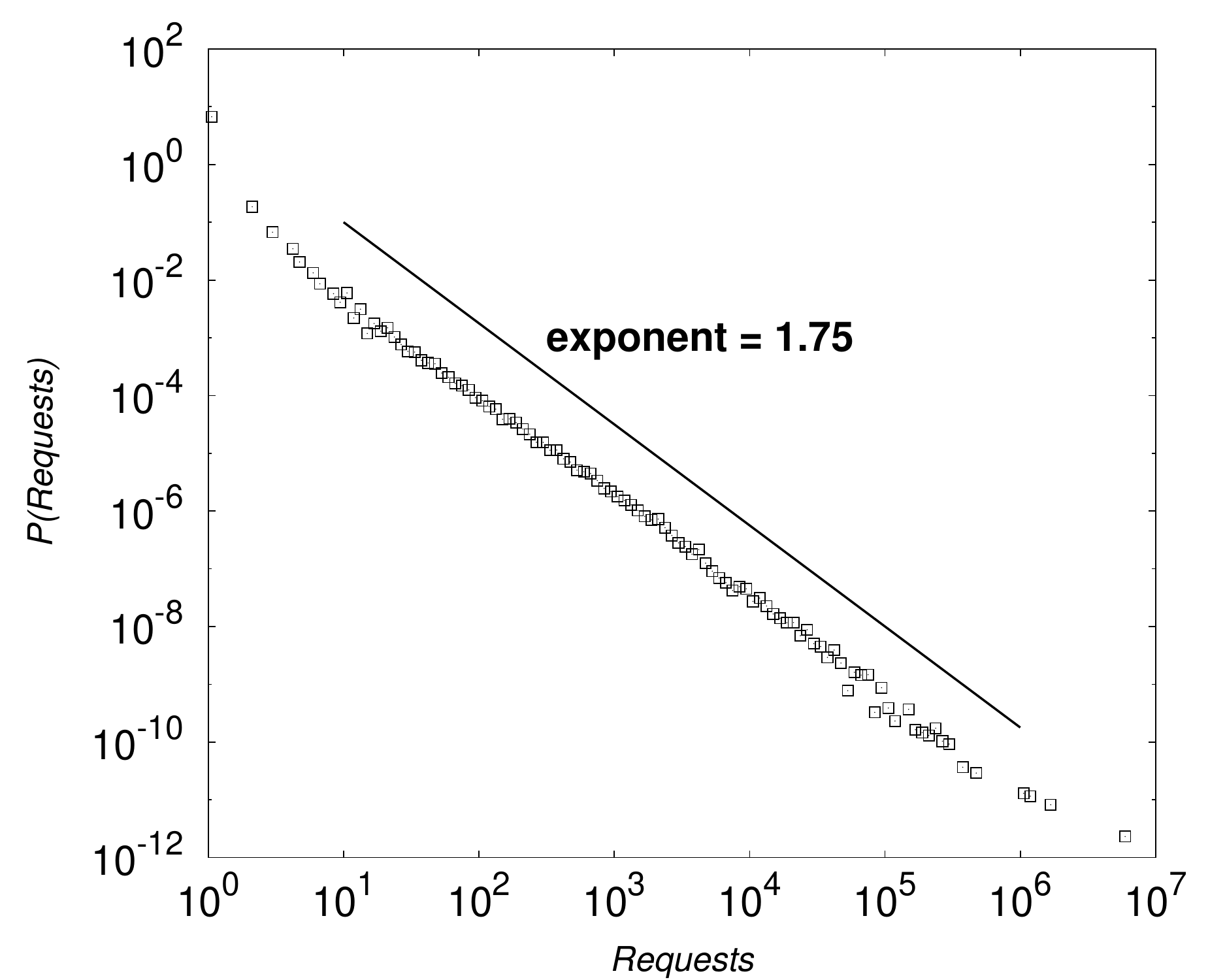} \\
\raisebox{1.8in}{B} & \includegraphics[width=0.75\columnwidth]{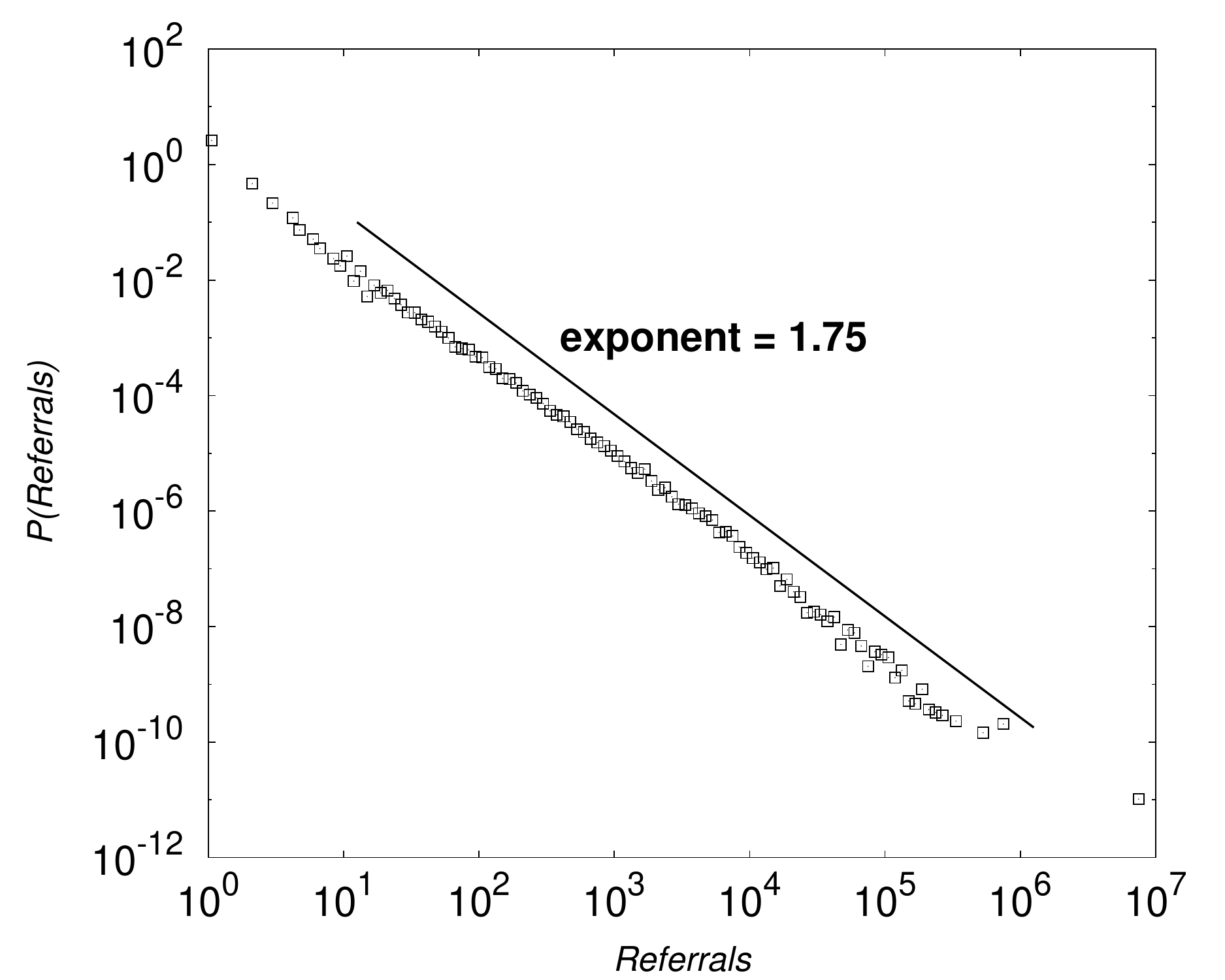}
\end{tabular}
\end{center}
\caption{
  Distributions of in-strength (A) and out-strength (B) for each Web
  server in the data set.  In these and the following plots in this
  paper, power-law distributions are fitted by least-squares
  regression on log values with log-bin-averaging and verified using
  the cumulative distributions and maximum likelihood
  methods~\protect\cite{Clauset07powerlaws}.}
\label{fig:host_request}
\end{figure}

Our first priority in analyzing this data set was to verify that its
statistics were consistent with those of previous studies.  A previous
study performed by several of the authors used a similar data
collection method to perform completely anonymized click records from
the entire Indiana University community of roughly 100,000
users~\cite{meiss-2008-wsdm}.  In that study, we found that the
distribution of the number of requests directed to each Web server
(``in-strength'', or $s_{in}$) could be well fitted by a power law
$\Pr(s_{in}) \sim s_{in}^{-\gamma}$ with exponent $\gamma \approx
1.8$.  The distribution of $s_{in}$ in the present data set is
consistent with this, as shown in Figure~\ref{fig:host_request}A; the
distribution is linear on a log-log scale for nearly six orders of
magnitude with a slope of roughly 1.75.  Similarly, in the previous
study, we found that the number of requests citing each Web server as
a referrer (``out-strength'', or $s_{out}$) could be approximated by a
power law with $\gamma \approx 1.7$.  In the current study, we find
that $\gamma \approx 1.75$ for $s_{out}$, as shown in
Figure~\ref{fig:host_request}B.  The overall distribution of traffic
is thus found to be in concordance with previous results.

\begin{figure}[tb]
\begin{center}
\begin{tabular}[t]{cc}
\raisebox{1.8in}{A} & \includegraphics[width=0.75\columnwidth]{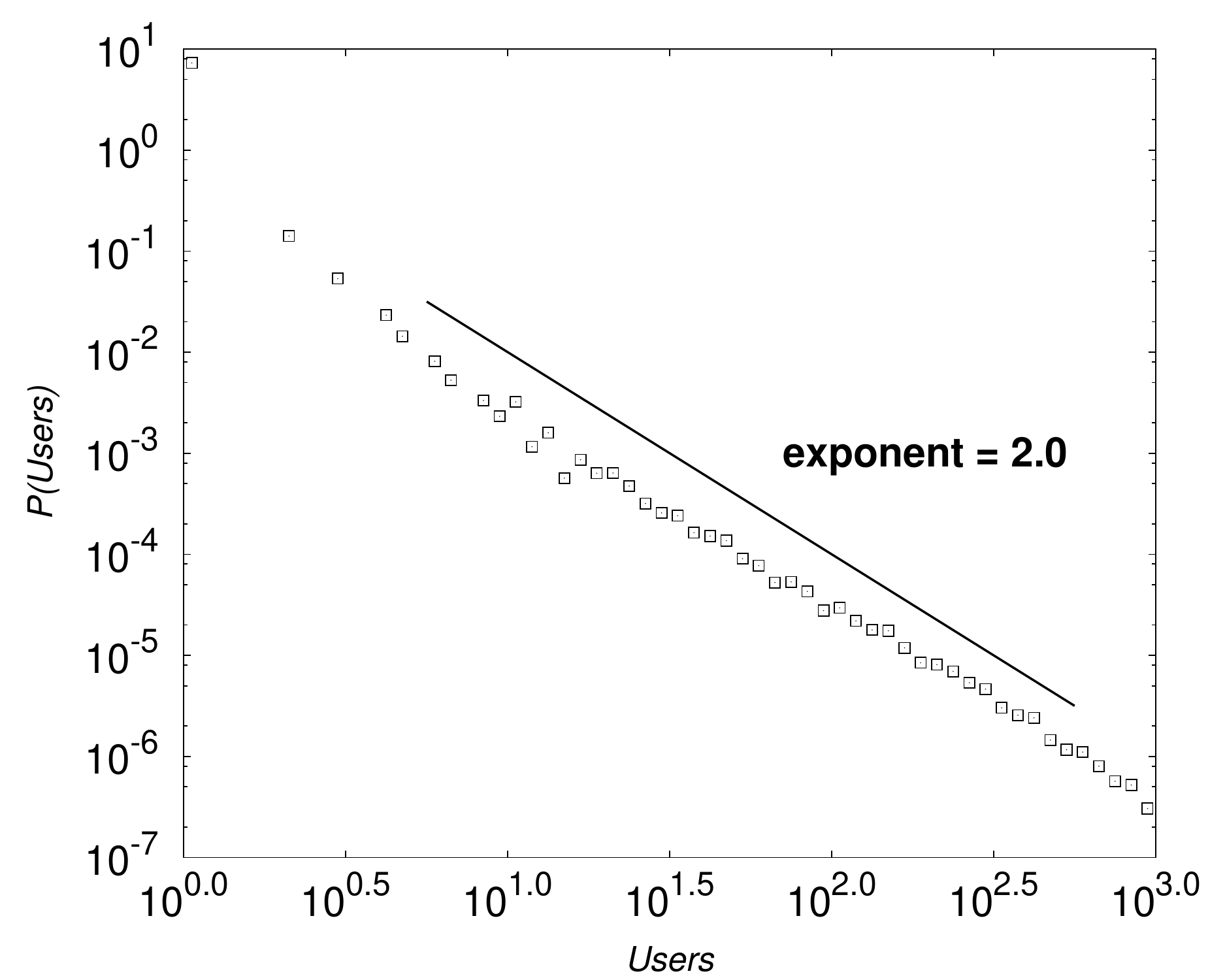} \\
\raisebox{1.8in}{B} & \includegraphics[width=0.75\columnwidth]{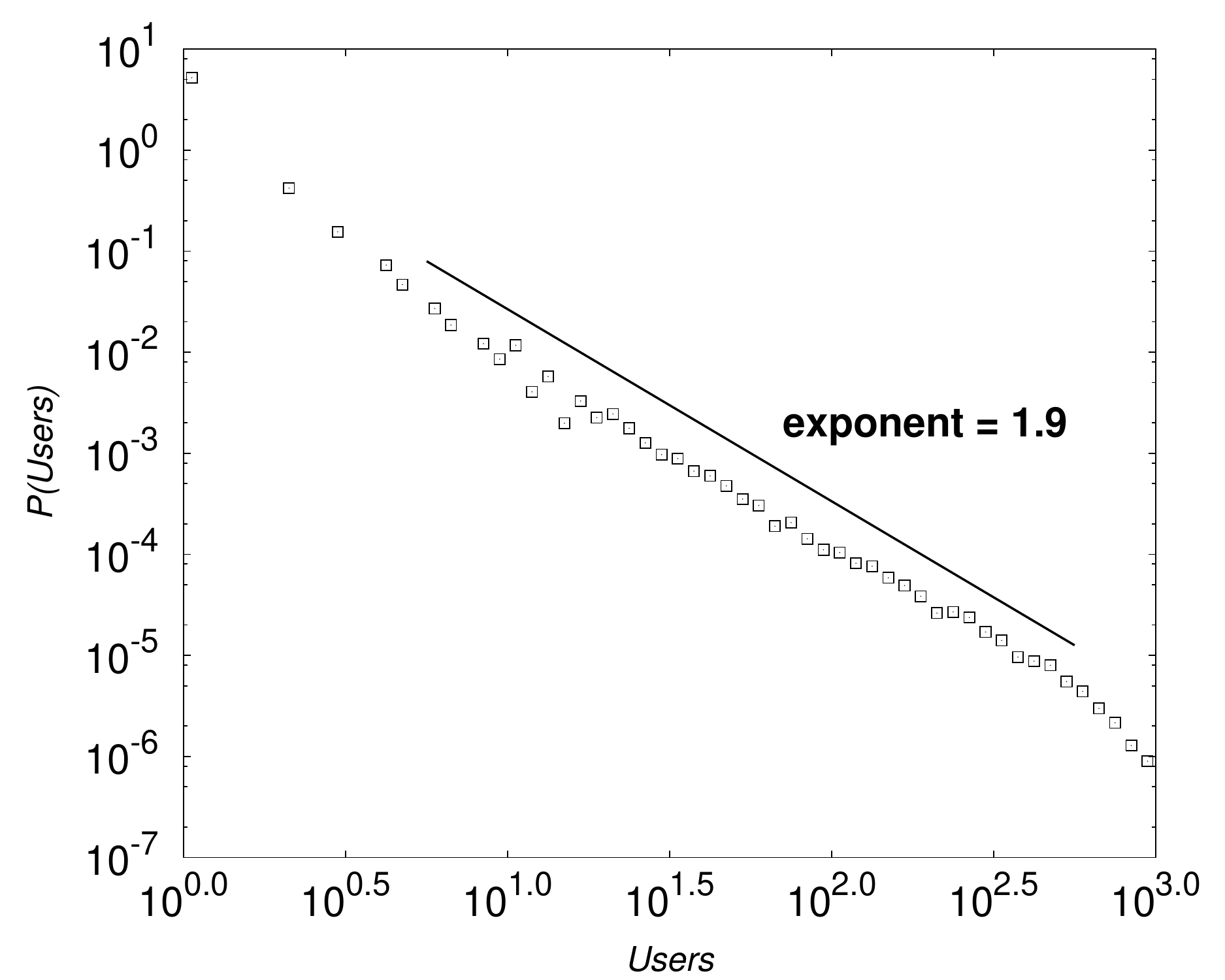}
\end{tabular}
\end{center}
\caption{
  Distributions of the number of unique users incoming to (A) and
  outgoing from (B) for each Web server in the data set.  These
  distributions serve as a rough measure of the popularity of a Web
  site and imply that the potential audience of the most popular sites
  is essentially unbounded.}
\label{fig:users_per_host}
\end{figure}

The previous study was conducted under conditions of complete
anonymity for users, retaining not even information as to whether two
requests came from the same or different clients.  Because the present
data set does attribute each request to a particular user, we were now
able to examine the relative popularity of Web server as measured by
the number of distinct users contributing to their traffic.  As shown
in Figures~\ref{fig:users_per_host}A and \ref{fig:users_per_host}B, we
find that the distribution of the number of users $u$ contributing to
the inbound traffic of a Web server is well approximated by a power
law $\Pr(u) \sim u^{-\beta}$ with $\beta \approx 2.0$ and the outbound
by a power law with $\beta \approx 1.9$.

These exponents require further comment.  Because $\beta~<~3$, the
variance diverges as the distribution grows and is bounded only by the
finite size of the data collection.  Furthermore, if $\beta \leq 2$,
as seems to be the case for both incoming and outgoing popularity, the
distributions lack any intrinsic mean as well.  This implies the lack
of any inherent ceiling of popularity for Web sites, regardless of the
size of the user population.  Indeed, the data show that the social
networking site Facebook is a popular destination for almost 100\% of
the students in our study, handily eclipsing any major search engine
or news site.

%------------------------------------------------------------------------
% (5) User-based properties
%------------------------------------------------------------------------

\section{User-based properties}
\label{section-user}

The behavior of individual users is of critical interest for not only
models of traffic, but also applications such as network anomaly
detection and the design of anonymization tools.  Because nearly all
per-server distributions in Web traffic exhibit scale-free properties
and have extremely heavy tails, one might anticipate that the same
would be true of Web users.  If the statistics that describe user
behavior lack well-defined central tendencies, than very little
individual behavior can be described as anomalous.  However, since any
given user has only finite time to devote to Web surfing, we know that
user-based distributions must be bounded.  The question is whether we
can characterize ``normal'' individual traffic.  If we can establish a
clear picture of the typical user, unusual users are easy to identify
and have a more difficult time maintaining their anonymity.

\begin{figure}[tb]
\begin{center}
\begin{tabular}[t]{cc}
\raisebox{1.8in}{A} & \includegraphics[width=0.75\columnwidth]{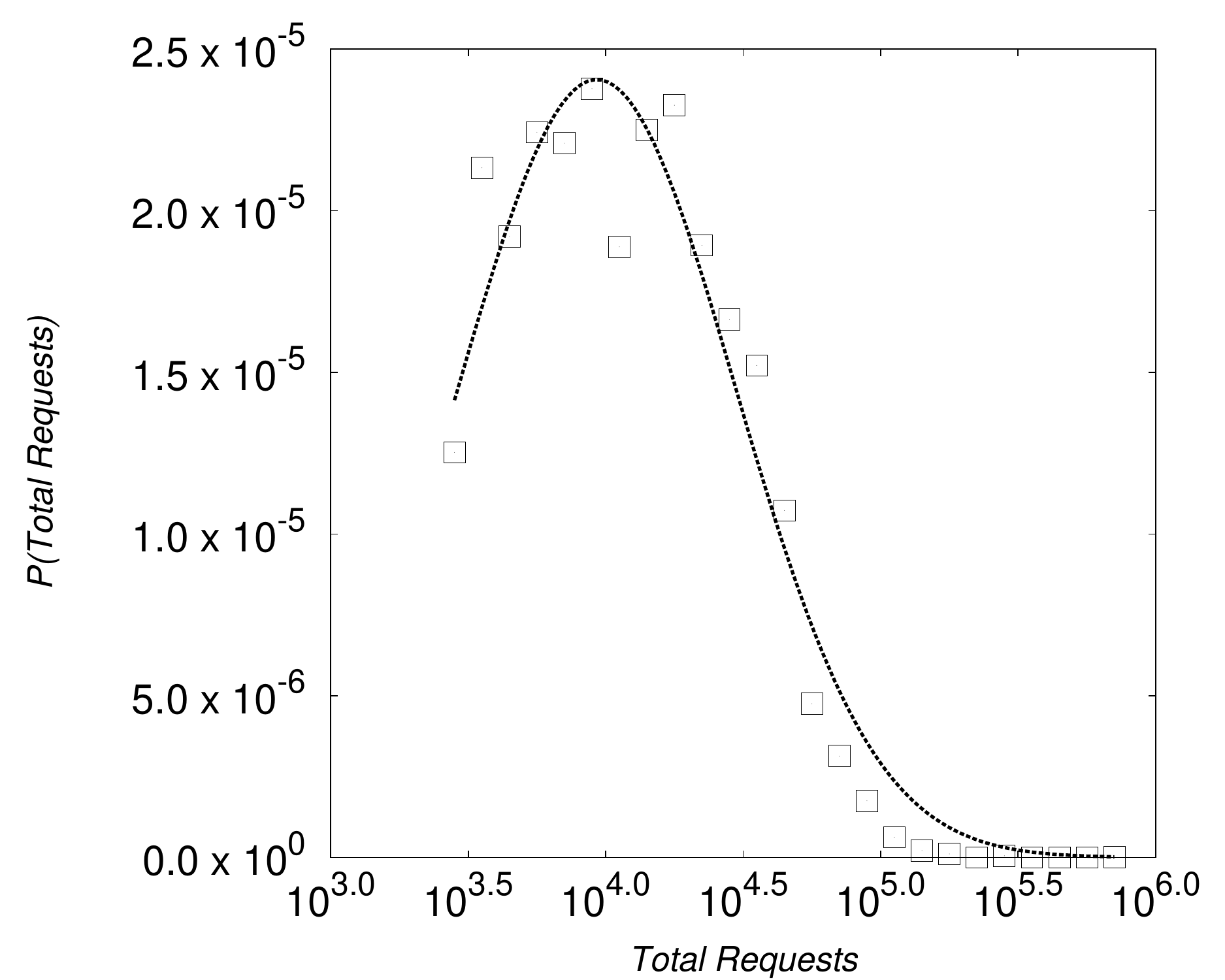} \\
\raisebox{1.8in}{B} & \includegraphics[width=0.75\columnwidth]{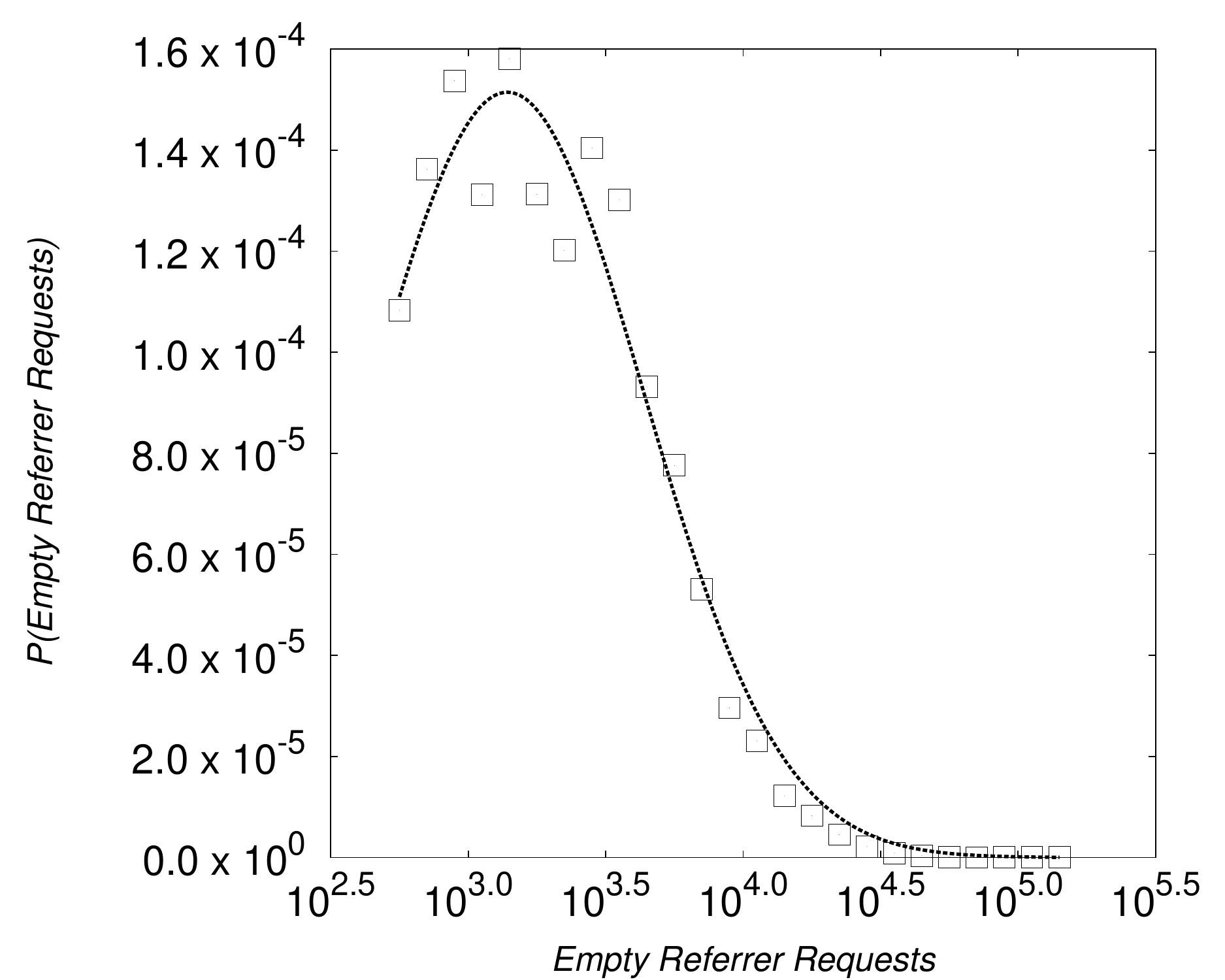}
\end{tabular}
\end{center}
\caption{
  Distributions of the number of requests made per user (A) and the
  number of empty-referrer requests made per user (B).  We show a
  reference log-normal fits to each distribution, which omit some
  low-traffic users as described in the text.}
\label{fig:requests_per_user}
\end{figure}

We first consider the distribution of sizes of the users' individual
click streams, both in terms of the total number of requests generated
by each user and the number of empty-referrer requests generated.  The
second distribution is of interest because it describes the number of
times a user has jumped directly to a specific page (e.g., using a
bookmark, start page, hyperlink in an e-mail, etc.) instead of
navigating there from already viewed pages.  The resulting
distributions are shown in Figures~\ref{fig:requests_per_user}A and
\ref{fig:requests_per_user}B.  Although the smaller size of this
distributions makes fitting more difficult, we do observe reasonably
strong log-normal fits for these distributions, finding that the
average user generated around 16,600 requests from 2,500 start pages
over the course of two months.  We removed from both distributions a
small number of users (roughly 50 in each case) whose click streams
were very small: under 2,500 requests or under 500 start pages.  Most
of these were users who did not begin generating any traffic until
late in the study, possibly because of new hardware or the approach of
final exams.

\begin{figure}[tb]
\begin{center}
\includegraphics[width=0.80\columnwidth]{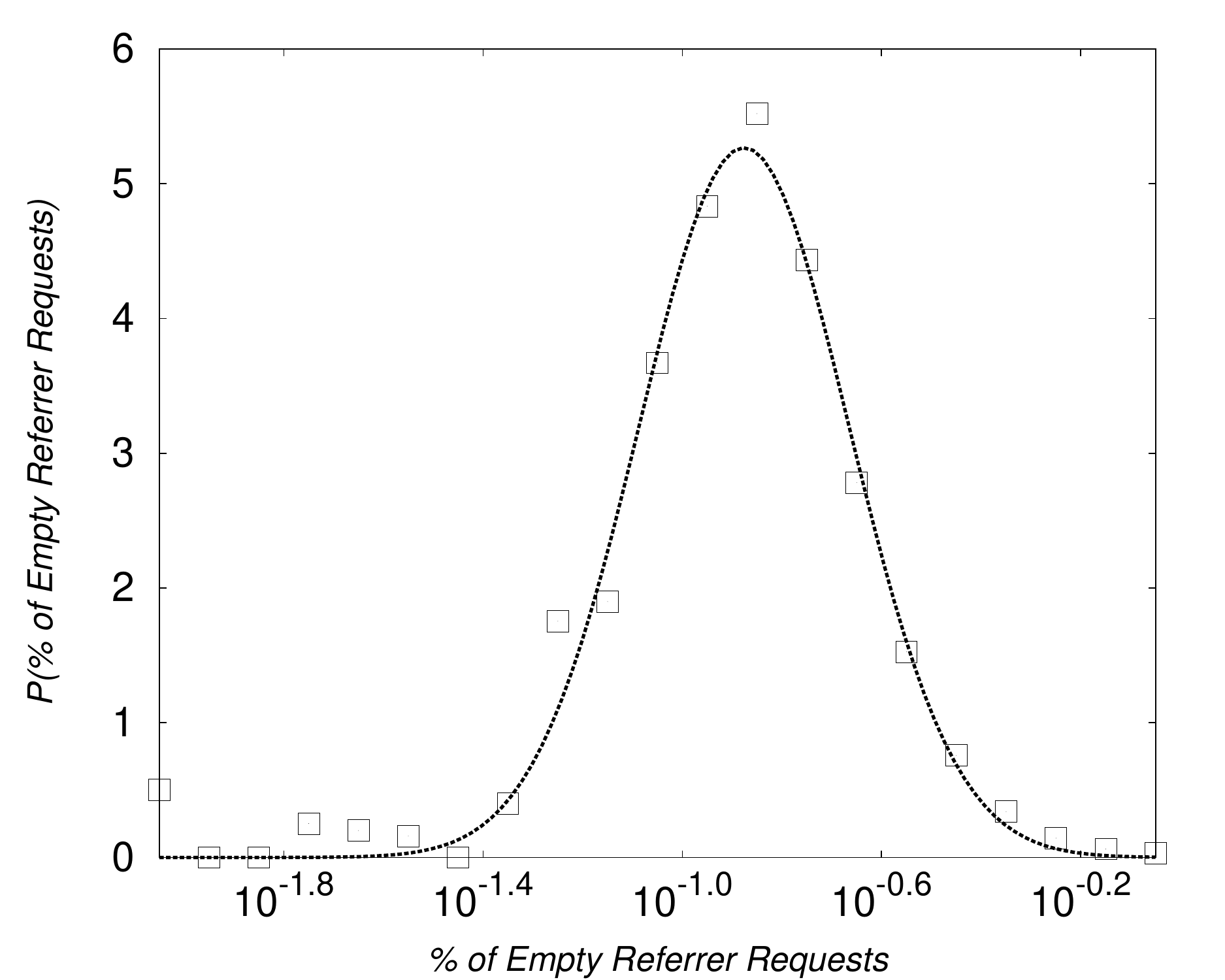}
\end{center}
\caption{
  Distribution of the proportion of empty-referrer requests made by
  each user, which roughly corresponds to the chance that a user will
  jump to a new location instead of continuing to surf.  We show a
  reference log-normal fit to the distribution.}
\label{fig:empty_proportion}
\end{figure}

We next examine the distribution of the ratio of the number of
empty-referrer requests to the total number of requests for each user.
This is a rough measure of the ``jump percentage'' (sometimes referred
to as the teleportation parameter) in the surfing behavior of users,
which is a value of critical importance to the PageRank
algorithm~\cite{Page98}.  A strong central tendency would imply that a
random surfer has a fairly constant jump probability
\textit{overall} even if the chance of jumping varies strongly from
page to page.  As shown in Figure~\ref{fig:empty_proportion}, we do
observe a strong fit to a log-normal distribution with a mean of about
15\%, which matches remarkably well the jump probability most often
used in Page\-Rank calculations.

\begin{figure}[tb]
\begin{center}
\includegraphics[width=0.75\columnwidth]{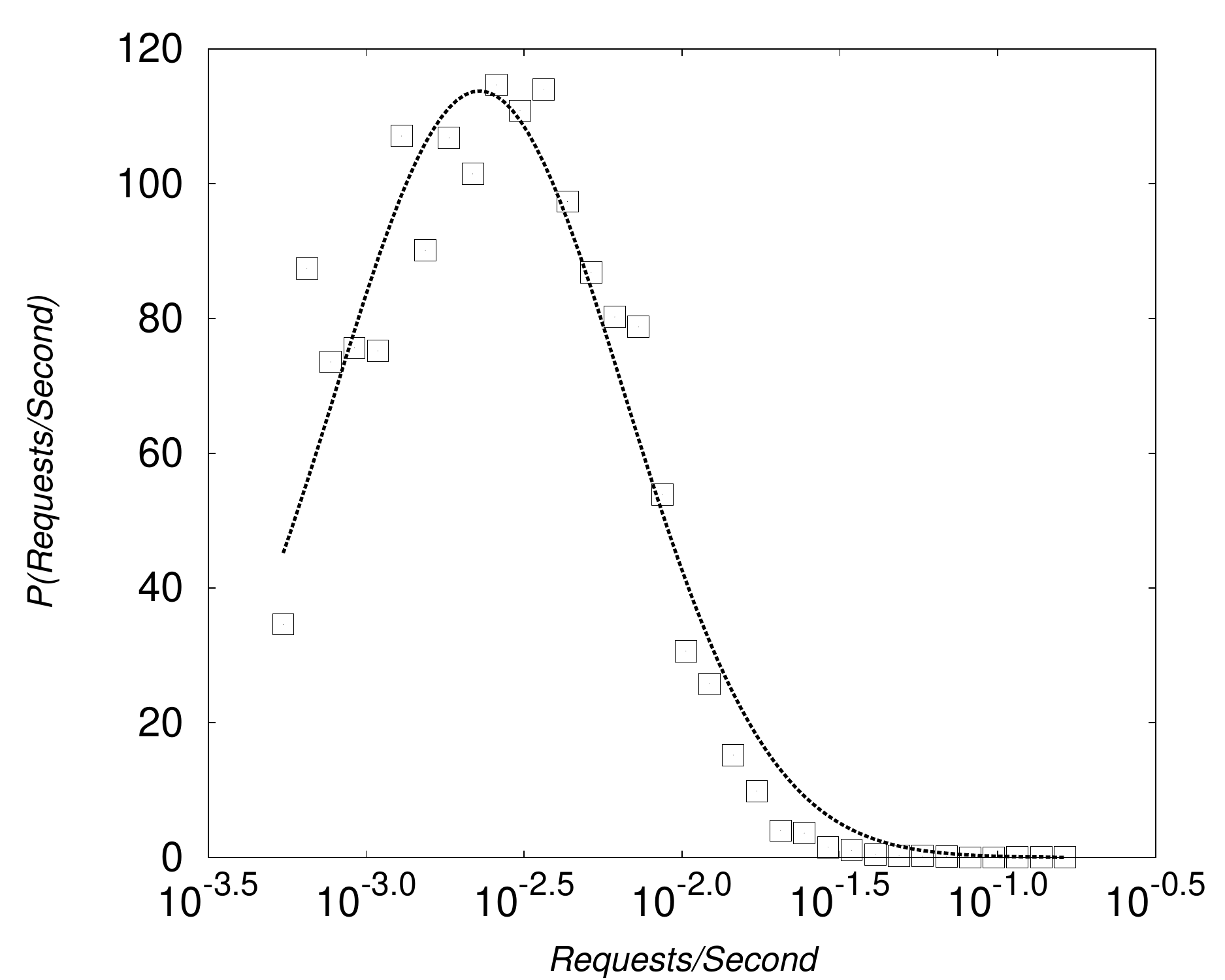}
\end{center}
\caption{
  Distribution of the number of requests per second made by each user,
  together with a reference log-normal fit to the distribution.}
\label{fig:request_per_sec}
\end{figure}

Besides the number of requests generated by each user, it is
interesting to inspect the rate at which those requests are generated.
Because not every user was active for the full duration of data
collection, this cannot be deduced directly from the distribution of
total requests.  In Figure~\ref{fig:request_per_sec} we show the
distribution of the number of requests per second for each user
generating an average of at least fifty requests over the time they
were active.  We again obtain a reasonable fit to a log-normal
distribution with a mean of about 0.0037 requests per second or 320
requests per day.

\begin{figure}[tb]
\begin{center}
\includegraphics[width=0.75\columnwidth]{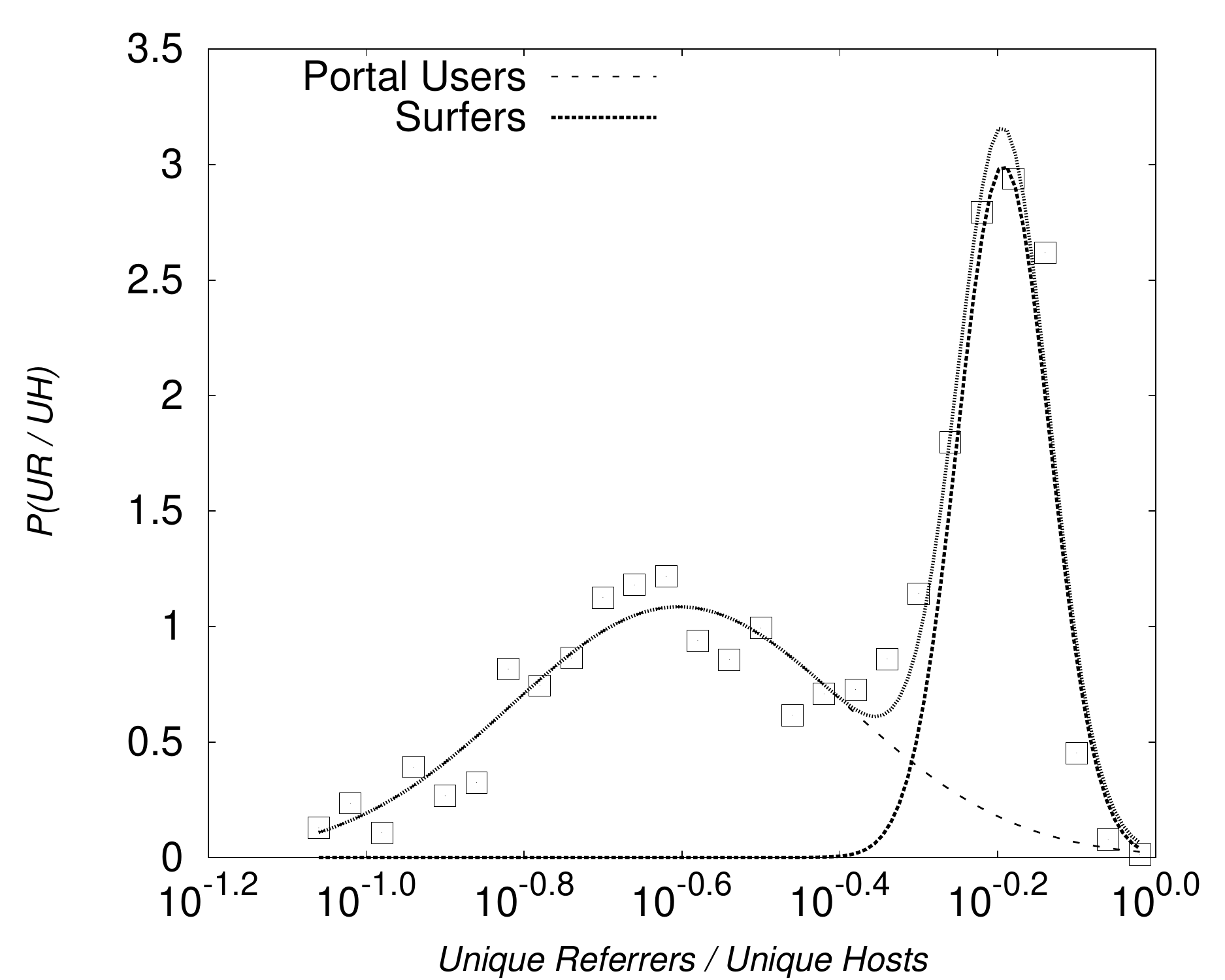}
\end{center}
\caption{
  Distribution of the ratio of unique referring sites to unique target
  sites for each user.  We approximate this bimodal distribution with
  two log-normals with means at 0.28 and 0.65.}
\label{fig:ref_per_host}
\end{figure}

Finally, we consider the ratio of the number of unique referring sites
to the number of unique target sites for each user.  This ratio serves
as a rough measure of the extent to which a user's behavior is
typified by searching or surfing.  If the number of referring hosts is
low as compared to the number of servers contacted, this implies that
the user browses the Web through a fairly small number of gateway
sites, such as search engines, social networking sites, or a personal
bookmark file.  If the number of referring hosts is high compared to
the number of servers contacted, this implies that a user is more
given to surfing behavior: they discover new sites through navigation.
We observe in Figure~\ref{fig:ref_per_host} that the distribution is
bimodal, implying the existence of two groups of user: one more
oriented toward portals and one more oriented toward browsing.  Portal
users visit on average almost four sites for each referrer, while
surfers visit only about 1.5 sites.  In support of this
characterization, we note that the overall mean ratio is 0.54, but
that this drops to 0.37 among users with more than 60\% of their
traffic connected to Facebook in some way.

%------------------------------------------------------------------------
% (6) Session properties
%------------------------------------------------------------------------

\section{Defining sessions}
\label{section-session}

When we contemplate the design of Web applications or modeling the
behavior of Web users, we are naturally drawn to the notion of a Web
session.  The constrained environments in which we most often observe
users on the Web make it easy to imagine that a user sits down at the
computer, fires up a Web browser, issues a series of HTTP requests,
then closes the browser and moves on to other, unrelated tasks.  This
is certainly the behavior we observe when users visit a research lab
to participate in a study or must dial into a modem pool before
beginning to surf the Web.  The subjects of the present study did not
fit these conditions; they have 24-hour access to dedicated network
connections, and we observed the traffic they generate in an
environment that is both their home and their workplace.  This
distinction made us suspect that we might face some difficulty in
selecting the optimal value for a timeout.

In our first attempt to segment individual click streams into
sessions, we settled on a five-minute inactivity timeout as a
reasonable start, a decision informed by previous research in the
field~\cite{Cho05WebDB}.  We found that each user's click stream split
into an average of 520 sessions over the two-month period.  A typical
session lasted for a bit over ten minutes and included around sixty
requests to twelve different Web servers.  These values seemed
plausible for the population: one can imagine the typical student
participating in ten ten-minute Web sessions every day.

The straightforward approach of identifying sessions using inactivity
timeouts thus seemed promising, so we experimented with a variety of
different timeouts to find an optimal value.  Because of the
log-normal distributions of user activity we had seen, it did
not seem unreasonable to suppose that some of the per-session
statistics would remain relatively constant as we adjusted the timeout
and others would show dramatic changes in the neighborhood of some
critical threshold.

\begin{figure}[tb]
\begin{center}
\includegraphics[width=\columnwidth]{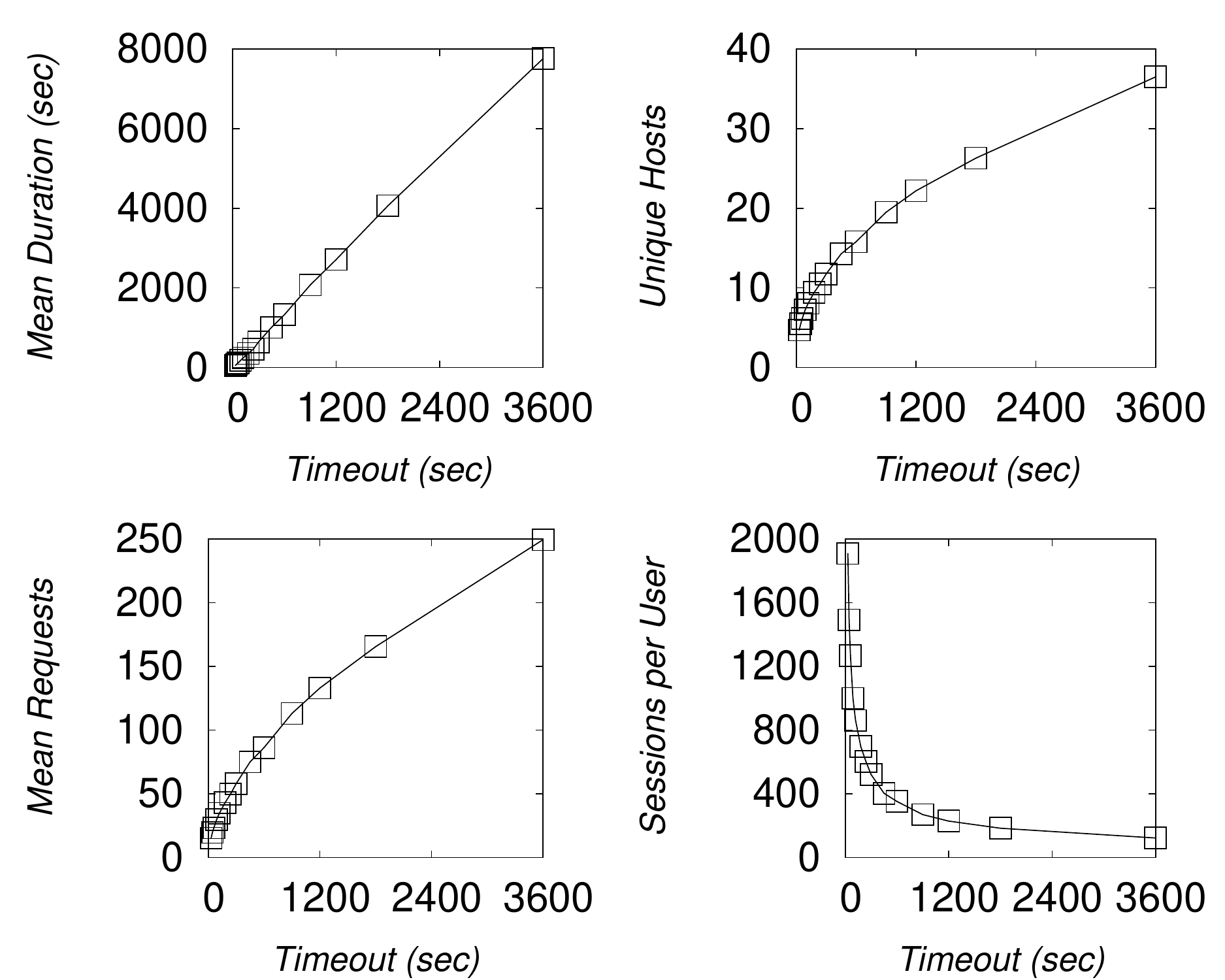}
\end{center}
\caption{
  Session statistics as a function of timeout.  Top left: Mean
  duration of sessions in seconds.  Top right: Mean number of hosts
  contacted.  Bottom left: Mean number of requests.  Bottom right:
  Mean number of sessions per user.}
\label{fig:session_stat}
\end{figure}

The results, shown in Figure~\ref{fig:session_stat}, show this to be
far from the case.  All the statistics we examined (mean number of
sessions per user, session duration, number of requests, and
number of hosts contacted) turn out to exhibit strong and regular
dependence on the particular timeout used.  They have no large
discontinuities, areas of particular stability, or even local maxima
and minima.  This implies there is no reason based on the data to
select one timeout threshold over any other; the choice is purely
arbitrary and becomes the prime determiner of all relevant statistics.

While we did expect some dependence on the timeout value, this result
surprised us.  We conjectured that the observed behavior might be a
side-effect of considering every user's sessions as part of the same
distribution; if we were to observe the click streams of individual
users, we might see more pronounced clustering of HTTP requests in
time.

\begin{figure}[tb]
\begin{center}
\includegraphics[width=\columnwidth]{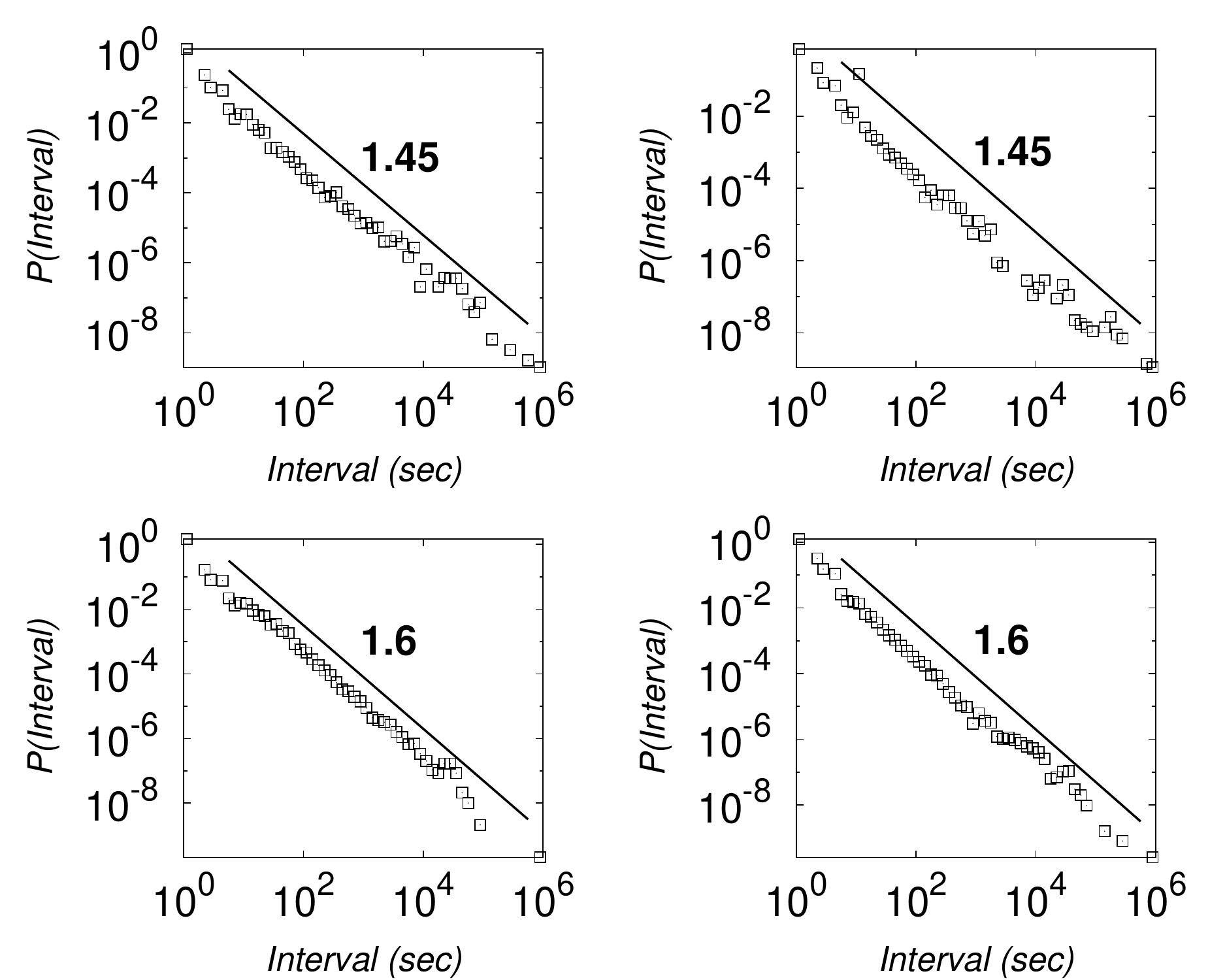}
\end{center}
\caption{
  The distributions of the time between successive requests for four
  randomly selected users.  Each distribution is well approximated by
  power laws with exponents below two, suggesting both unbounded
  variance and the lack of a well-defined mean.}
\label{fig:user_interclick}
\end{figure}

To test this notion, we picked several users at random.  For each one,
we found the distribution of ``interclick times'', defined as the
number of seconds elapsed between each pair of successive page
requests.  A user with $n$ requests in their click stream would thus
yield $n - 1$ interclick times.  If a user's activity were typified by
tight bursts of requests with long periods of inactivity between them,
we would expect to find a steep decline in the tail of the probability
density function at the point where the interclick time no longer
represents the time between requests but the time between entire
sessions.  Instead, we found that the users' distributions of
interclick times could be closely approximated by power-law
distributions $\Pr(\Delta t) \sim \Delta t^{-\tau}$ over nearly six
orders of magnitude, as shown in Figure~\ref{fig:user_interclick}.
Moreover, we found $\tau < 2$ in each case, suggesting that there is
no central tendency at all to the time between a user's requests, and
that no delay can really be considered atypical of a user.

\begin{figure}[tb]
\begin{center}
\includegraphics[width=0.75\columnwidth]{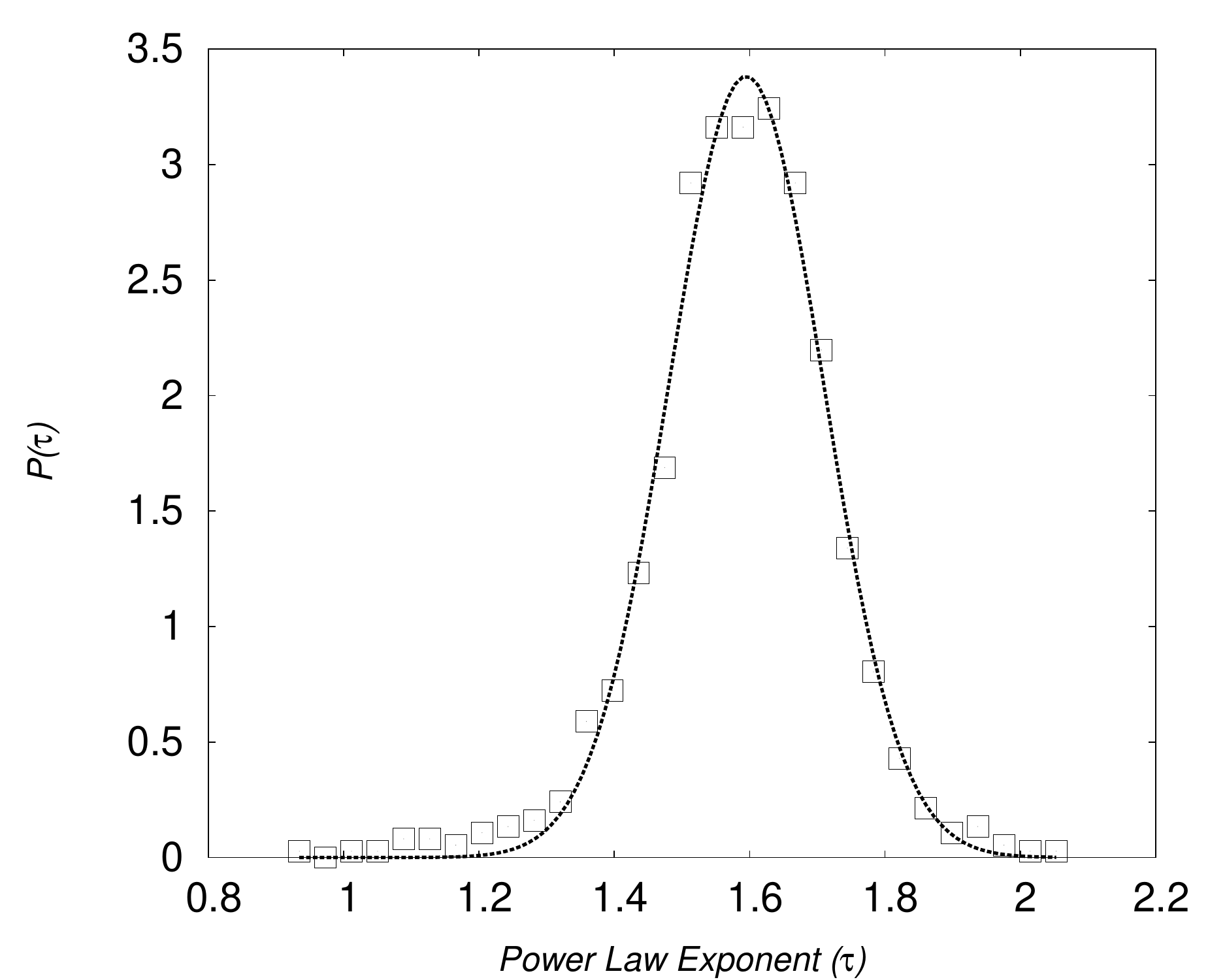}
\end{center}
\caption{
  The distribution of the exponent $\tau$ for the best power-law
  approximation to the distribution of interclick times for each user.
  The fit is a normal distribution with mean $\langle \tau \rangle =
  1.6$ and standard deviation $\sigma = 0.1$.  These low values of
  $\tau$ indicate unbounded variance and the lack of any central
  tendency in the time between requests for a Web surfer.}
\label{fig:gamma}
\end{figure}

The possibility remained that these randomly selected users might be
outliers, so we automated the process of calculating the probability
density function for a user's distribution of interclick times using
log-binned histograms and then fitting the result to a power law
approximation.  We were able to fit each distribution to a power law
with a mean $R^2$ value of 0.989.  The resulting distribution of
power-law exponents, shown in Figure~\ref{fig:gamma}, is strongly
normal with a mean value $\langle \tau \rangle \approx 1.6$.  This
confirmed the finding that interclick times have no central tendency;
in fact, scale-free behavior is so pervasive that a user agent
exhibiting \textit{regularity} in the timing of its requests would
constitute an anomaly.  These results make it clear that a robust
definition of Web session cannot be based on a simple timeout.

The next natural approach would be to segment a click stream into
sessions based on the rolling average of the number of clicks per unit
time dropping below a threshold value.  This approach turns out to be
even more problematic than using a simple timeout, as there are now
two parameter to consider: the width of the window for the rolling
average and the threshold value.  If the window selected is too
narrow, then the rolling average will often drop all the way to zero,
and the scheme becomes prey to the same problems as the simple
timeout.  If the window selected is too large, then the rolling
average is so insensitive to change that meaningful segmentation is
impossible.  In the end, the choice is once again arbitrary.
Moreover, examination of the moving average click rate for several
users shows that the magnitudes of the spikes in the click rate fit a
smooth distribution.  This makes the selection of a threshold value
arbitrary as well: the number of sessions becomes a function of the
threshold rather than a feature of the data itself.

Logging both the referring URL and target URL for HTTP requests makes
possible a third and more robust approach to constructing user
sessions.  We expand on the notion of \textit{referrer trees} as
described in \cite{Cho05WebDB} to segment a user's click stream
into a set of \textit{logical sessions} based on the following
algorithm:

\begin{enumerate}
\item{Initialize the set of sessions $T$ and the map $N : U \mapsto T$ from URLs to sessions.}
\item{For each request with referrer $r$ and target $u$:
  \begin{enumerate}
  \item{If $r$ is the empty referrer, create a new session $t$ with root node $u$, and set $N(u) = t$.}
  \item{Otherwise, if the session $N(r)$ is well-defined, attach $u$ to $N(r)$ if necessary, and set $N(u) = N(r)$.}
  \item{Otherwise, create a new session $t$ with root node $r$ and single leaf node $u$, and set $N(r) = N(u) = t$.}
  \end{enumerate}
}
\end{enumerate}

This algorithm assembles requests into sessions based on the referring
URL of a request matching the target URL of a previous request.
Requests are assigned to the session with the most recent use of the
referring URL.  Furthermore, each instance of a request with an empty
referrer generates a new logical session.

Before we examine the properties of the logical sessions defined by
this algorithm, we must highlight the differences between logical
sessions and our intuitive notion of session.  A logical session does
not represent a period of time in which a user opens a Web browser,
browses some set of Web sites, and then leaves the computer.  It
instead connects requests related to the same browsing behavior.  If
the user opens links in multiple tabs or uses the browser's back
button, the subsequent requests will all be part of the same logical
session.  If the user then jumps directly to a search engine, they
will start a new logical session.  Tabbed browsing and use of the back
button make it entirely possible for a user to have multiple active
logical session at any point in time.  A user who always keeps a
popular news site open in a browser tab might have a logical session
related to that site that lasts for the entire collection period.
Logical sessions thus segment a user's click stream into
\textit{tasks} rather than time intervals.  They also enjoy the
advantage of being defined without reference to any external
parameters, making their properties comparable across data sets and
insensitive to the judgment of individual researchers.

\begin{figure}[tb]
\begin{center}
\begin{tabular}[t]{cc}
\raisebox{1.8in}{A} & \includegraphics[width=0.75\columnwidth]{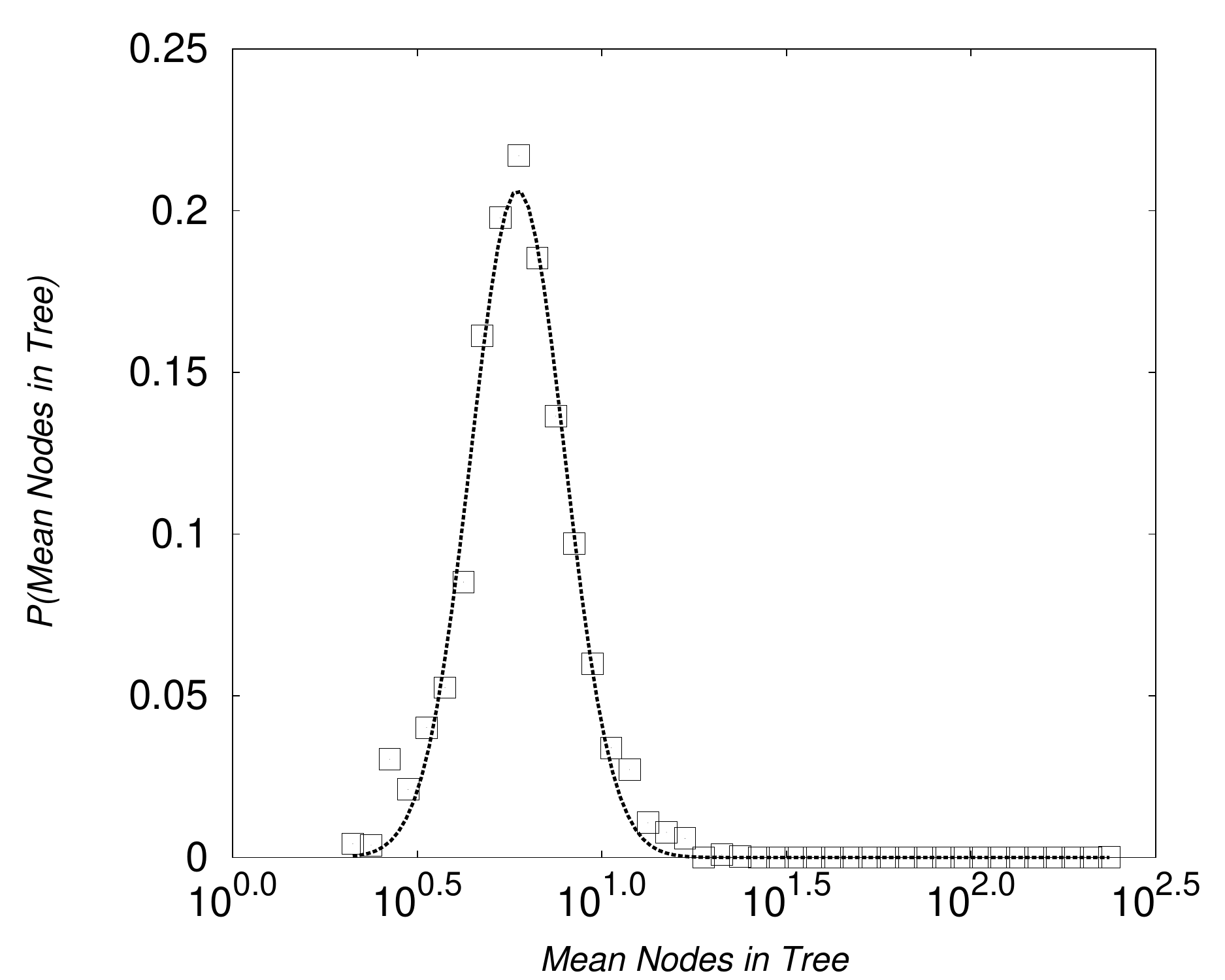} \\
\raisebox{1.8in}{B} & \includegraphics[width=0.75\columnwidth]{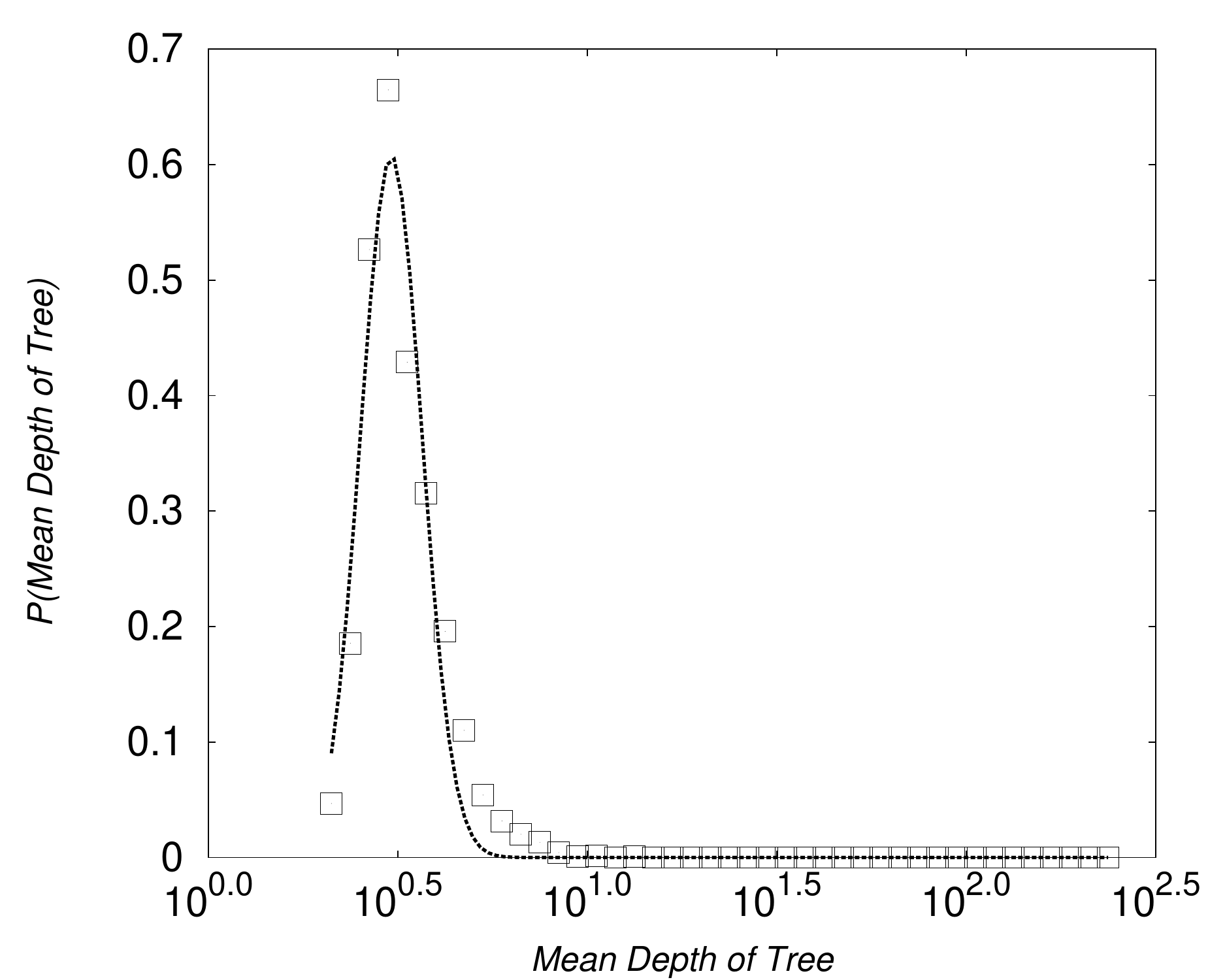}
\end{tabular}
\end{center}
\caption{
  Distributions of the mean number of requests per logical session per
  user (A) and the mean depth of logical session per user (B).
  In both cases, we consider only non-trivial trees.  We show
  reference log-normal fits to each distribution.}
\label{fig:tree_per_user}
\end{figure}

The first statistics of interest for these logical sessions concern
their tree structures.  The number of nodes in the tree is a count of
the number of requests in the session.  In
Figure~\ref{fig:tree_per_user}A, we show the probability density
function for the per-user distribution of the average size of a
logical session.  This distribution is well approximated with a
log-normal function, showing that the typical user has a mean of
around 6.1 requests per session.  The depth of the trees indicates the  % was 5.6, updated after tree fix -MRM
extent to which users follow chains of links during a Web-related
task.  Figure~\ref{fig:tree_per_user}B shows the distribution of the
average depth of the logical sessions for each user.  It is again
well-approximated with a log-normal, showing that a typical user's
sessions have a depth of about three links.  In other words, an
average session usually travels no more than two clicks away from the
page on which it began.

\begin{figure}[tb]
\begin{center}
\includegraphics[width=0.75\columnwidth]{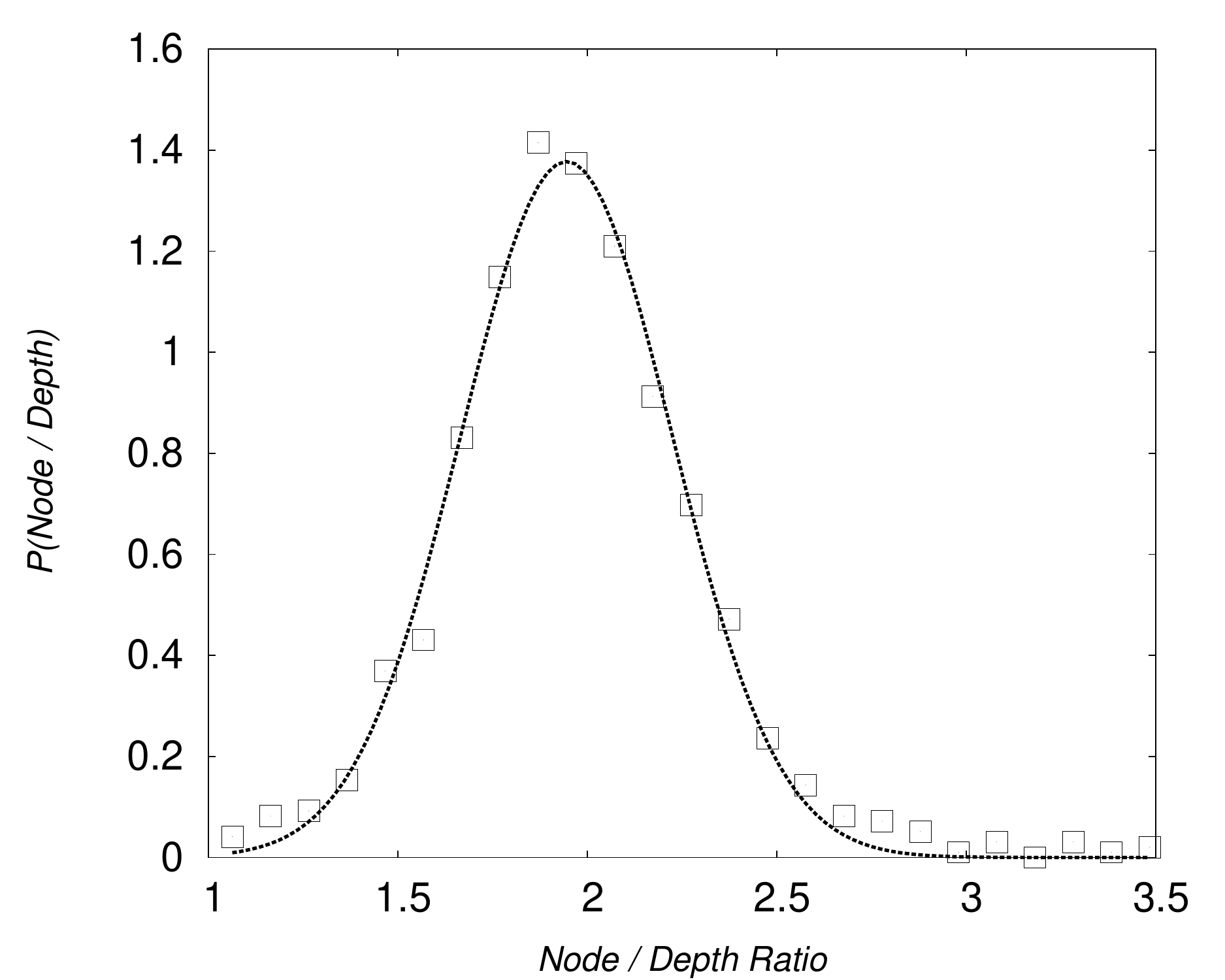}
\end{center}
\caption{
  The distribution of the average ratio of the node count to the tree
  depth for the logical sessions of each user.  The fit is a normal
  distribution with mean $\mu = 1.94$ and standard deviation $\sigma =
  0.28$, showing that the branching factor of logical sessions is
  significantly greater than one.}
\label{fig:tree_node_over_depth}
\end{figure}

The ratio between the number of nodes in each tree and its depth is
also of interest.  If this ratio is equal to 1, then the tree is just
a sequence of clicks, which corresponds to the assumptions of the
random walker model used by PageRank.  As this ratio grows past 1, the
branching factor of the tree increases, the assumptions of PageRank
break down, and a random walker must either backtrack or split into
multiple agents.  Figure~\ref{fig:tree_node_over_depth} shows the
distribution of the average node-to-depth ratio for each user, which
is well-approximated by a normal distribution with mean $\langle \mu
\rangle = 1.94$.  We thus see that sessions have structure that cannot be  % was 1.83, updated after tree fix -MRM
accurately modeled by a stateless random walker: there must be
provision for backtracking or branching.

\begin{figure}[tb]
\begin{center}
\begin{tabular}[t]{cc}
\raisebox{1.8in}{A} & \includegraphics[width=0.75\columnwidth]{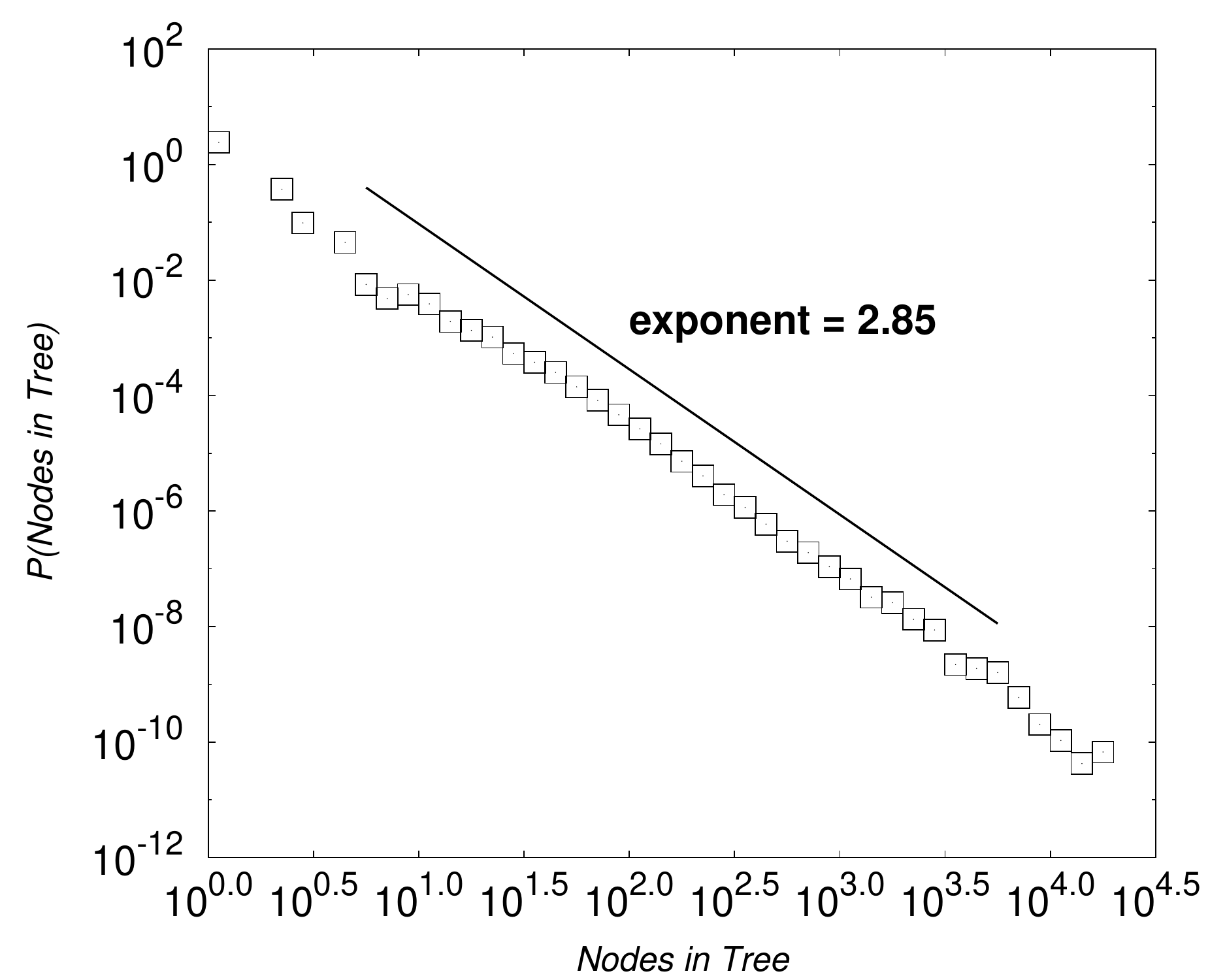} \\
\raisebox{1.8in}{B} & \includegraphics[width=0.75\columnwidth]{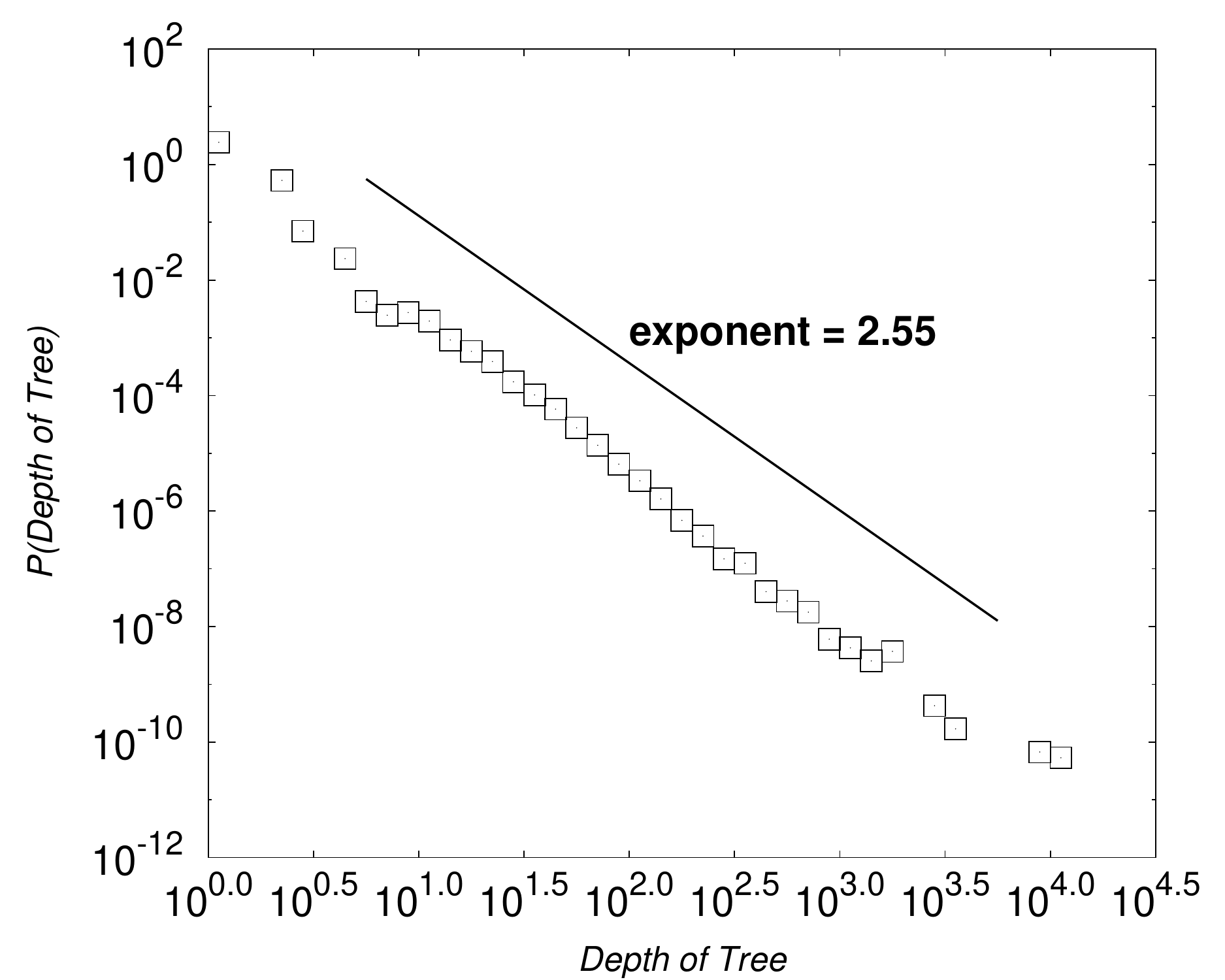}
\end{tabular}
\end{center}
\caption{
  Distributions of the number of requests per logical session (A) and
  the depth of each logical session (B), with reference power-law
  fits.}
\label{fig:tree_count}
\end{figure}

Although there is a strong central tendency to the mean size and depth
of a logical session for each user, the same does not hold for logical
sessions in general.  In Figure~\ref{fig:tree_count}, we show the
distributions of the node count and depth for logical sessions
aggregated across all users.  When we remove per-user identifying
information in this way, we are once again confronted with
heavy-tailed distributions exhibiting unbounded variance.  This
implies that the detection of automated browsing traffic is a much
more tractable task if some form of client identity can be retained.

\begin{figure}[tb]
\begin{center}
\includegraphics[width=0.75\columnwidth]{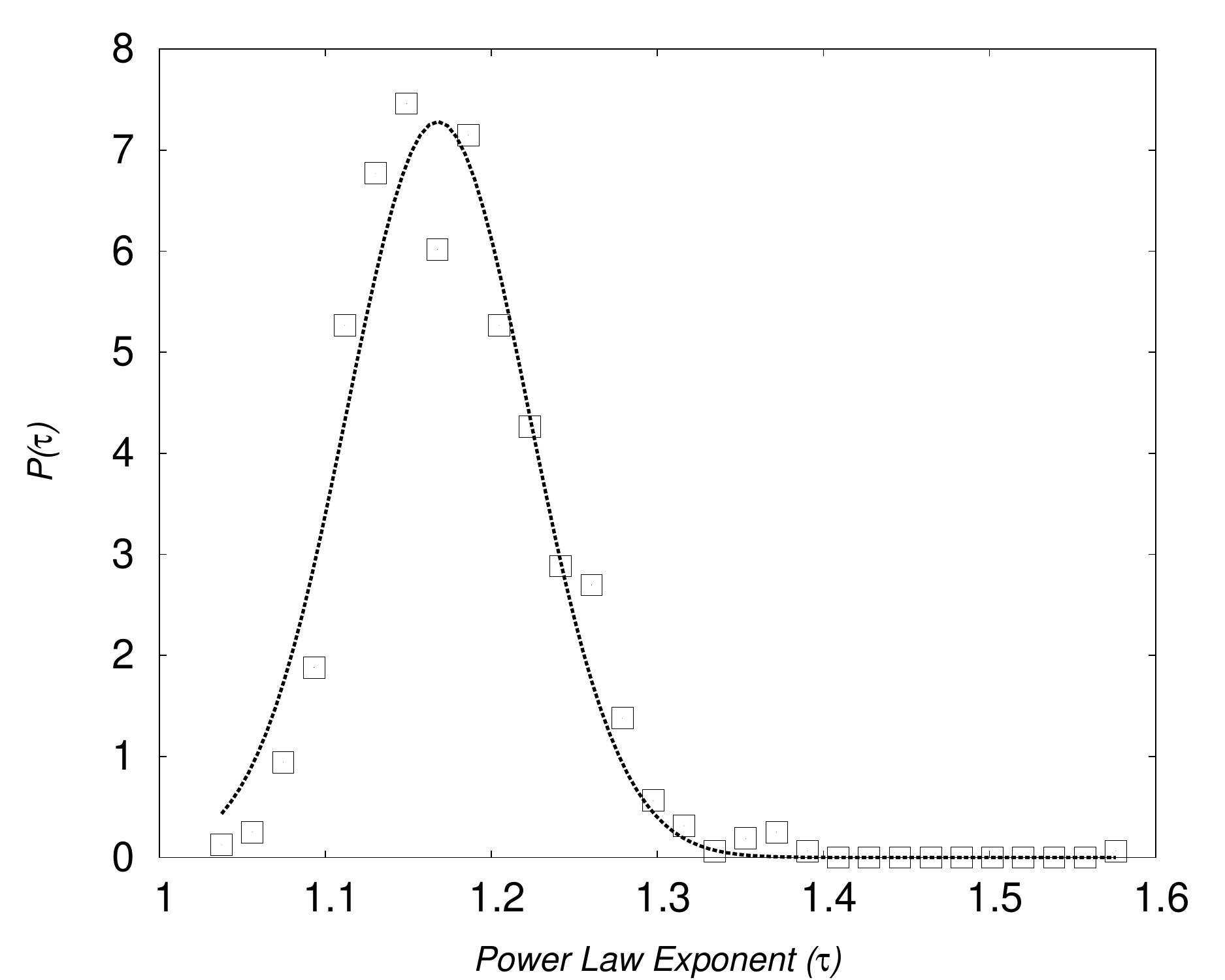}
\end{center}
\caption{
  The distribution of exponent $\tau$ for the best power-law
  approximation to the distribution of logical session duration for
  each user.  The fit is a normal distribution with mean $\langle \tau
  \rangle = 1.2$ and standard deviation $\sigma = 0.06$.  These low
  values of $\tau$ indicate unbounded variance and the lack of any
  central tendency in the duration of a logical session.}
\label{fig:bracket_gamma}
\end{figure}

Even though we have defined sessions logically, they can still be
considered from the perspective of time.  If we calculate the
difference between the timestamp of the request that first created the
session and the timestamp of the most recent request to add a leaf
node, we obtain the duration of the logical session.  When we examine
the distribution of the durations of the sessions of a user, we
encounter the same situation as for the case of interclick times:
power-law distributions $\Pr(\Delta t) \sim \Delta t^{-\tau}$ for
every user.  Furthermore, when we consider the exponent of the best
power-law fit of each user's data, we find the values are normally
distributed with a mean value $\langle \tau \rangle \approx 1.2$, as
shown in Figure~\ref{fig:bracket_gamma}.  No user has a well-defined
mean duration for their logical sessions; as also suggested by the
statistics of interclick times, the presence of strong regularity in a
user's session behavior would be anomalous.

It is natural to speculate that we can get the best of both worlds by
extending the definition of a logical session to include a timeout, as
was done in previous work on referrer trees~\cite{Cho05WebDB}.  Such a
change is quite straightforward to implement: we simply modify the
algorithm so that a request cannot attach to an existing session
unless the attachment point was itself added within the timeout.  This
allows us to have one branch of the browsing tree time out while still
allowing attachment on a more active branch.

\begin{figure}[tb]
\begin{center}
\includegraphics[width=\columnwidth]{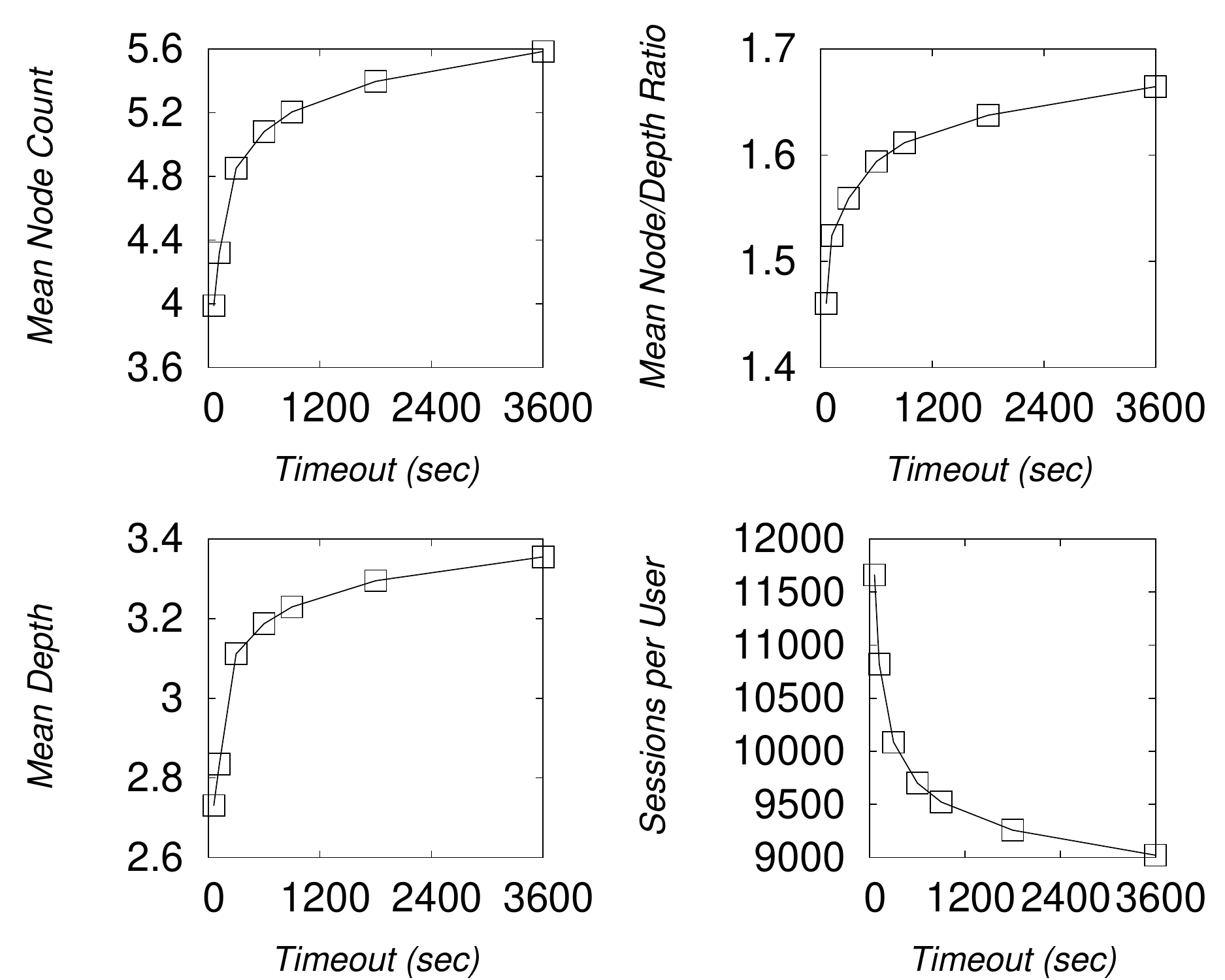}
\end{center}
\caption{
  Logical session statistics as a function of timeout.  Top left: Mean
  node count.  Top right: Mean tree depth.  Bottom left: Mean ratio of
  node count to tree depth.  Bottom right: Mean number of sessions per
  user.}
\label{fig:logical_stat}
\end{figure}

While the idea is reasonable, we unfortunately find that the addition
of such a timeout mechanism once again makes the statistics of the
sessions strongly dependent on the particular timeout selected.  As
shown in Figure~\ref{fig:logical_stat}, the number of sessions per
user, mean node count, mean depth, and ratio of nodes to tree depth
are all dependent on the timeout.  On the other hand, in contrast to
sessions defined purely by timeout, this dependence becomes smaller as
the timeout increases, suggesting that logical sessions with a timeout
of around 15 minutes may be a reasonable compromise for modeling and
further analysis.

%------------------------------------------------------------------------
% (7) Conclusions
%------------------------------------------------------------------------

\section{Conclusions}

In this paper we have built on the network-sniffing approach to
gathering Web traffic that we first explored
in~\cite{meiss-2008-wsdm}, extending it to track the behavior of
individual users.  The resulting data set provides an unprecedented
and accurate picture of human browsing behavior in a hypertext
information space as manifested by over a thousand undergraduate
students in their residences.

The data confirm previous findings about long-tailed distributions in
site traffic and reveal that the popularity of sites is likewise
unbounded and without any central tendency.  They also show that while
many aspects of Web traffic have been shown to obey power laws, these
power-law distributions often represent the aggregate of distributions
that are actually log-normal at the user level.  The lack of any
regularity in interclick times for Web users leads to the conclusion
that sessions cannot be meaningfully defined with a simple timeout,
leading to our presentation of logical sessions and an algorithm for
deriving them from a click stream.  These logical sessions illustrate
further drawbacks of the random surfer model and can be modified to
incorporate timeouts in a relatively robust way.

These findings have direct bearing on future work in modeling user
behavior in hypertext navigation.  The stability of the proportion of
empty-referrer requests across all users implies that although not
every page is equally likely to be the cause of a jump, the overall
chance of a jump occurring is constant in the long run.  The finding
that the branching factor of the logical sessions is definitely
greater than one means that plausible agent-based models for random
walks must incorporate state, either through backtracking or
branching~\cite{bouklit-2005-backrank}.

Our indications as to which distributions show central tendencies and
which do not are of critical importance for anomaly detection and
anonymization.  To appear plausibly human, an agent must not stray too
far from the expected rate of requests, proportion of empty-referrer
requests, referrer-to-host ratio, and node count and tree depth values
for logical sessions.  Because these are log-normal distributions,
agents cannot deviate more than a multiplicative factor away from
their means.  At the same time, a clever agent must mimic the
heavy-tailed distributions of the spacing between requests and
duration of sessions; too \textit{much} regularity appears artificial.

Although our method of collection does afford us with a large volume
of data, it suffers from several disadvantages which we are working to
overcome in future studies.  First, our use of the file extension (if
any) in requested URLs is a noisy indicator of whether a request truly
represent a page fetch.  We are also unable to detect whether any
request is actually satisfied or not; many of the requests may
actually result in server errors or redirects.  Both of these problems
could be largely mitigated without much overhead by capturing the
first packet of the server's response, which should indicate an HTTP
response code and a content type in the case of successful requests.

This data set is inspiring the development of an agent-based model
that replaces the uniform distributions of Page\-Rank with more
realistic distributions and incorporates bookmarking behavior to
capture the branching behavior observed in logical
sessions~\cite{goncalves-2008-remembering}.

%------------------------------------------------------------------------
% (8) Acknowledgments
%------------------------------------------------------------------------

\section*{Acknowledgments}

The authors would like to thank the Advanced Network Management
Laboratory at Indiana University and Dr.~Jean Camp of the IU School of
Informatics for support and infrastructure.  We also thank the network
engineers of Indiana University for their support in deploying and
managing the data collection system.  Special thanks are due to
Alessandro Flammini for his insight and support during the analysis of
this data.

This work was produced in part with support from the Institute for
Information Infrastructure Protection research program.  The I3P is
managed by Dartmouth College and supported under Award Number
2003-TK-TX-0003 from the U.S.~DHS, Science and Technology Directorate.
This material is based upon work supported by the National Science
Foundation under award number 0705676.  This work was supported in
part by a gift from Google.  Opinions, findings, conclusions,
recommendations or points of view in this document are those of the
authors and do not necessarily represent the official position of the
U.S.~Department of Homeland Security, Science and Technology
Directorate, I3P, National Science Foundation, Indiana University,
Google, or Dartmouth College.

\bibliographystyle{abbrv}

\begin{thebibliography}{}

\end{thebibliography}


\begin{thebibliography}{10}

\bibitem{agichtein-2006-user}
E.~Agichtein, E.~Brill, and S.~Dumais.
\newblock Improving {Web} search ranking by incorporating user behavior
  information.
\newblock In {\em Proc. 29th ACM SIGIR Conf.}, 2006.

\bibitem{borders-2004-webtap}
K.~Borders and A.~Prakash.
\newblock Web tap: Detecting covert web traffic.
\newblock In {\em In Proceedings of the 11th ACM Conference on Computer and
  Communication Security}, pages 110--120. ACM Press, 2004.

\bibitem{bouklit-2005-backrank}
M.~Bouklit and F.~Mathieu.
\newblock Backrank: an alternative for pagerank?
\newblock In {\em WWW '05: Special interest tracks and posters of the 14th
  international conference on World Wide Web}, pages 1122--1123, New York, NY,
  USA, 2005. ACM.

\bibitem{catledge95characterizing}
L.~D. Catledge and J.~E. Pitkow.
\newblock Characterizing browsing strategies in the {World-Wide Web}.
\newblock {\em Computer Networks and ISDN Systems}, 27(6):1065--1073, 1995.

\bibitem{Clauset07powerlaws}
A.~Clauset, C.~R. Shalizi, and M.~E.~J. Newman.
\newblock Power-law distributions in empirical data.
\newblock Technical report, arXiv:0706.1062v1 [physics.data-an], 2007.

\bibitem{cockburn-what}
A.~Cockburn and B.~McKenzie.
\newblock What do {Web} users do? {An} empirical analysis of {Web} use.
\newblock {\em Intl. Journal of Human-Computer Studies}, 54(6):903--922, 2001.

\bibitem{erman-2007-identifying}
J.~Erman, A.~Mahanti, M.~Arlitt, and C.~Williamson.
\newblock Identifying and discriminating between web and peer-to-peer traffic
  in the network core.
\newblock In {\em {WWW}}, pages 883--892, 2007.

\bibitem{Fortunato05egalitarian}
S.~Fortunato, A.~Flammini, F.~Menczer, and A.~Vespignani.
\newblock Topical interests and the mitigation of search engine bias.
\newblock {\em Proc. Natl. Acad. Sci. USA}, 103(34):12684--12689, 2006.

\bibitem{goncalves-2008-remembering}
B.~Gon{\c{c}}alves, M.~Meiss, J.~J. Ramasco, A.~Flammini, and F.~Menczer.
\newblock Remembering what we like: Toward an agent-based model of web traffic.
\newblock In {\em WSDM (Late-breaking papers)}, 2009.

\bibitem{goncalves08-2}
B.~Gon{\c{c}}alves and J.~J. Ramasco.
\newblock Human dynamics revealed through web analytics.
\newblock {\em Phys. Rev. E}, 78:026123, 2008.

\bibitem{liu-2008-browserank}
Y.~Liu, B.~Gao, T.~Y. Liu, Y.~Zhang, Z.~Ma, S.~He, and H.~Li.
\newblock {BrowseRank}: Letting web users vote for page importance.
\newblock In {\em {SIGIR}}, 2008.

\bibitem{luxenburger-2004-querylog}
J.~Luxenburger and G.~Weikum.
\newblock {\em Query-Log Based Authority Analysis for Web Information Search},
  volume 3306 of {\em Lecture Notes in Computer Science}, pages 90--101.
\newblock Springer Berlin / Heidelberg, 2004.

\bibitem{meiss-2008-wsdm}
M.~Meiss, F.~Menczer, S.~Fortunato, A.~Flammini, and A.~Vespignani.
\newblock Ranking {Web} sites with real user traffic.
\newblock In {\em {Proc. 1st {ACM} International Conference on Web Search and
  Data Mining (WSDM)}}, 2008.

\bibitem{Meiss05NetFlow}
M.~Meiss, F.~Menczer, and A.~Vespignani.
\newblock On the lack of typical behavior in the global {Web} traffic network.
\newblock In {\em Proc. 14th International World Wide Web Conference}, pages
  510--518, 2005.

\bibitem{meiss-2008-structural}
M.~Meiss, F.~Menczer, and A.~Vespignani.
\newblock Structural analysis of behavioral networks from the {I}nternet.
\newblock {\em Journal of Physics A}, 2008.

\bibitem{mobasher-2000-automatic}
B.~Mobasher, R.~Cooley, and J.~Srivastava.
\newblock Automatic personalization based on web usage mining.
\newblock {\em Communications of the ACM}, 43(8):141--151, 2000.

\bibitem{Page98}
L.~Page, S.~Brin, R.~Motwani, and T.~Winograd.
\newblock The {PageRank} citation ranking: {B}ringing order to the {Web}.
\newblock Technical report, Stanford University Database Group, 1998.

\bibitem{PandeyRoyOlstonChoChak05VLDB}
S.~Pandey, S.~Roy, C.~Olston, J.~Cho, and S.~Chakrabarti.
\newblock Shuffling a stacked deck: The case for partially randomized ranking
  of search engine results.
\newblock In K.~B{\"o}hm, C.~S. Jensen, L.~M. Haas, M.~L. Kersten, P.-{\AA}.
  Larson, and B.~C. Ooi, editors, {\em Proc. 31st International Conference on
  Very Large Databases (VLDB)}, pages 781--792, 2005.

\bibitem{Cho05WebDB}
F.~Qiu, Z.~Liu, and J.~Cho.
\newblock Analysis of user web traffic with a focus on search activities.
\newblock In A.~Doan, F.~Neven, R.~McCann, and G.~J. Bex, editors, {\em Proc.
  8th International Workshop on the Web and Databases (WebDB)}, pages 103--108,
  2005.

\bibitem{viecco-2009-privacy}
C.~Viecco, A.~Tsow, and L.~J. Camp.
\newblock Privacy-aware architecture for sharing web histories.
\newblock {\em {IBM} Systems Journal}, publication pending.

\bibitem{yang-2003-weblog}
Q.~Yang and H.~H. Zhang.
\newblock Web-log mining for predictive web caching.
\newblock {\em IEEE Trans. on Knowledge and Data Engineering},
  15(4):1050--1053, 2003.

\end{thebibliography}

\balancecolumns
\end{document}